# Flux pinning on nano-sized defects and critical current in HTc superconductors

*J. Sosnowski*
*Electrotechnical Institute*
*04-703 Warsaw, Pożaryskiego 28, Poland*

New results of the investigations critical current problems in the high temperature oxide superconductors are presented, basing on an analysis of the flux pinning on nano-sized centers. New model of the interaction pancake vortices with the nano-sized centers is proposed, while comparison of obtained results with others pinning forces approaches presented. The energy balance approach is applied, which leads to the appearance of the potential barrier determining the pancake vortices movement in the flux creep process. Various initial positions of the captured vortex have been analyzed, while variation of the screening currents for such configuration has been considered and initial calculations of the influence this effect on the current-voltage characteristics performed. The dependence of the critical current on the pinning centers parameters has been obtained while comparison with experimental data is given. Pinning force has been calculated and its temperature dependence, which is in qualitative agreement with an experiment. Critical current determines the magnetic induction profiles and therefore flux trapping, very important parameter from the point of view of an application HTc superconductors as permanent magnets. The influence of the critical current on the flux trapping in ceramic granular superconductors is calculated as well as influence of various other material parameters – grain radius, surface barrier, fulfillment of the ceramic with superconducting grains. Influence of the critical current on a.c. losses in second generation HTc tapes with magnetic substrate is discussed briefly too.

## 1. Introduction

The effect of capturing vortices in HTc superconductors has essential meaning from the point of view of electric current transport. Therefore analysis of this subject is important both from scientific as well as technical point of view. On relevance of this subject indicates a lot of papers devoted to the flux pinning (see for instance [1-4] and cited here references). Present paper from other side is the continuation of the previous author's works [5-8]. While large attention of the scientists is devoted mainly to the critical temperature analysis of the HTc superconductors, flux pinning in these materials also brings new unresolved yet problems in the comparison to the classic already low temperature superconductors. One of them is strong magnetic field decrease of the critical current in these materials, which in fact limits the real applications of HTc superconductors in the electric machines. From other side the layered structure of HTc materials leads to an existence of the pancake type vortices, which permits treat the pinning interaction as individual one between pinning center and vortex. We remind that in the classic 3D materials appear the flux lines multifold captured to the various centers, which requires to treat this case as the many body problem.

## 2. New model of the pinning interaction

High critical temperature oxide superconductors are characterized by unique superconducting parameters, as critical temperature but also critical current. An essential property of the high temperature oxide superconducting materials, such as critical current is very sensitive to the existence of structural defects. These defects called the columnar if created during the fast neutrons irradiation of the superconducting windings working in nuclear reactors are then of the nanometric size. They interact with the pancake vortices arising in HTc superconductors, stabilizing thus the vortex structure and permit in this way for the resistivity-less current flow. For description of this dynamic interaction new model of the vortex capturing has been developed, based on an analysis of the increase in the normal state energy during the pancake vortex deflection from the nano-sized pinning center, against its equilibrium position. Geometry of this interaction is shown in Fig. 2. Above approximation corresponds roughly to the consideration of the first two terms in the Ginzburg-Landau theory [9]. Two effects have been investigated: of the enhancement in the normal state energy during the vortex movement, as well as opposite one of the increase in the elasticity energy of the vortex lattice during the capturing process, deflecting vortex from its equilibrium position in the magnetic vortices array.

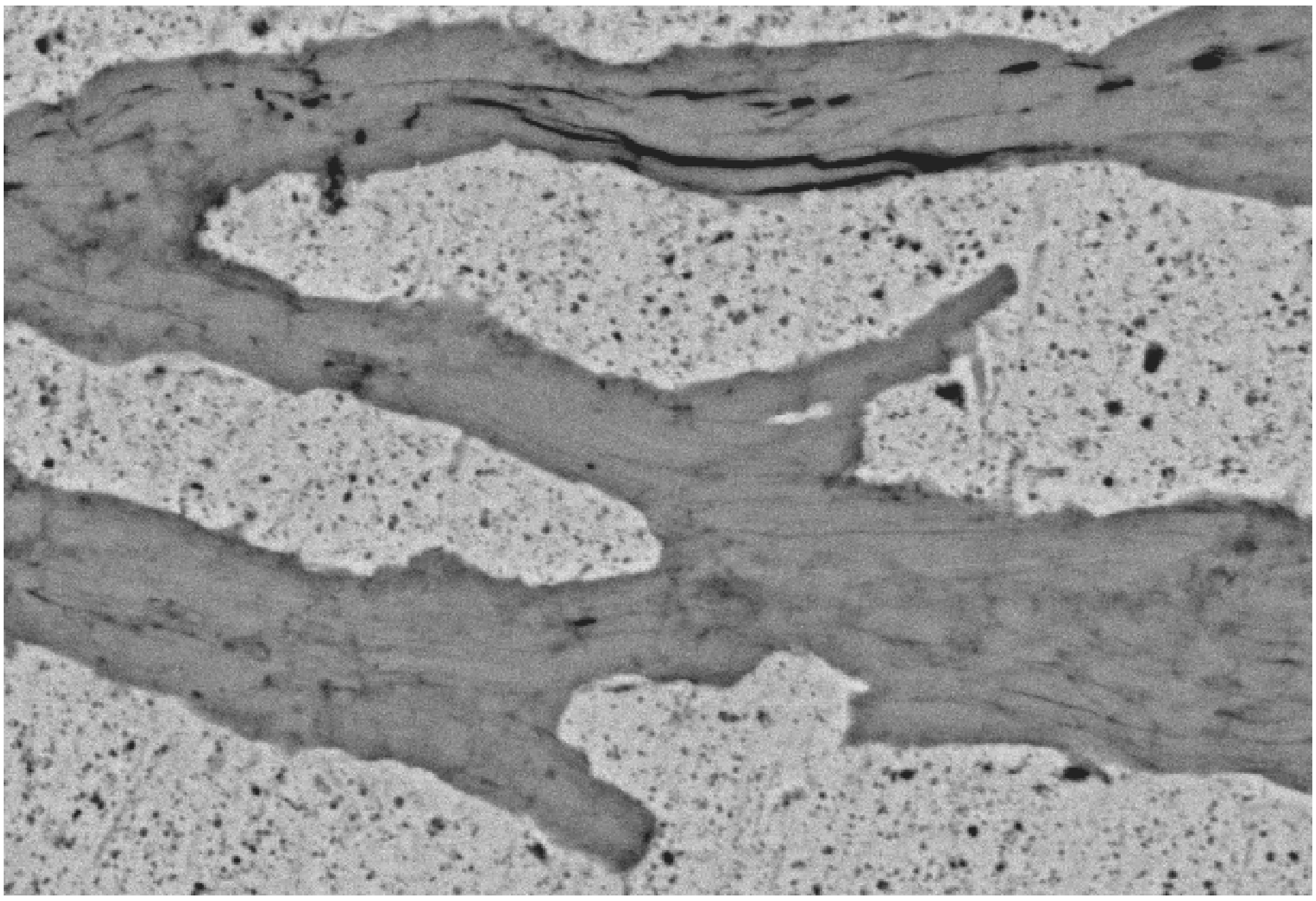

Fig. 1. Scanning electron microscopy (SEM) image of the cross-section, superconducting Bi-based tape, indicating the existence of the nano-structural defects (black points – among others nano-pores and $CuO_2$ grains). The length of the cross-section is 34 μm.

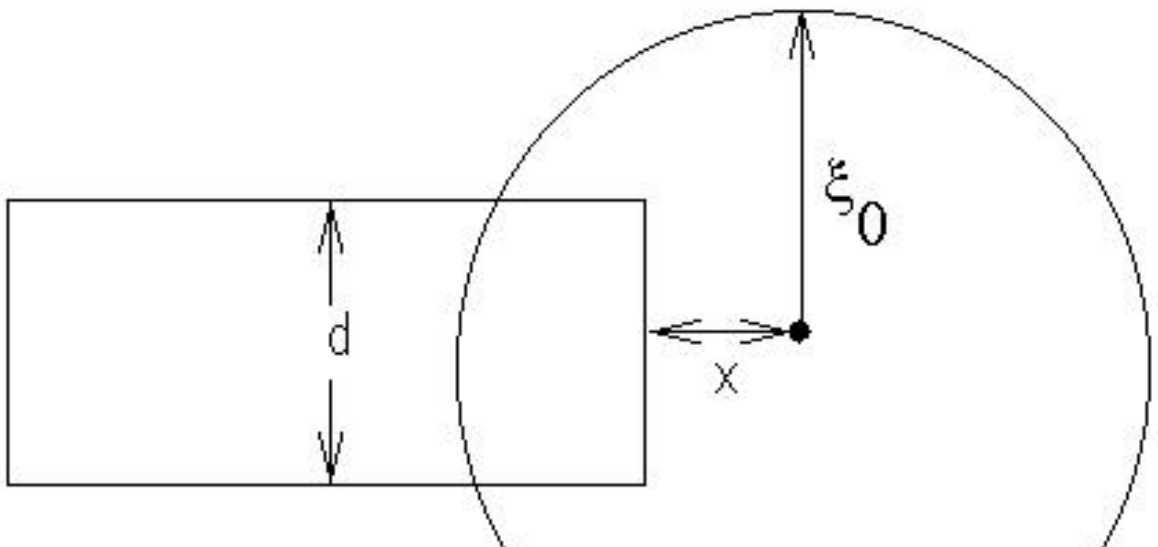

Fig. 2. View of the vortex core of the radius equal to the coherence length $\xi_0$ captured on the pinning center of the width ***d***

In the initial state of the captured vortex on the flat pinning center of nanosized dimensions ***d***, smaller than the coherence length ***ξ***, as is shown in the Fig. 2 for x = 0, the normal state energy of vortex core is:

$$U_1(0)=\frac{\mu_o H_c^2 l}{2}\left[\pi\xi^2-\xi^2\arcsin\frac{d}{2\xi}-\frac{d\xi}{2}\sqrt{1-\left(\frac{d}{2\xi}\right)^2}\right] \tag{1}$$

while for a vortex deflected by the distance ***x*** against this equilibrium position as shows Fig. 2, it is given as:

$$U_2(x)=\frac{\mu_o H_c^2 l}{2}\left[\pi\xi^2+dx-\xi^2\arcsin\frac{d}{2\xi}-\frac{d\xi}{2}\sqrt{1-\left(\frac{d}{2\xi}\right)^2}\right] \tag{2}$$

for $x<x_c$, where $\boldsymbol{x_C}$ is defined as:

$$x_c=\xi\sqrt{1-(\frac{d}{2\xi})^2} \tag{3}$$

The deflection of vortex is caused by the flow of the current in magnetic field, which leads to an arise of the Lorentz force tearing vortex from the initial captured position. For vortex deflected at distance larger than $\boldsymbol{x_C}$ normal energy connected with shifted vortex is given as follows:

$$U_3(x)=\frac{\mu_o H_c^2 l_p \xi^2}{2}\left[\frac{\pi}{2}+\arcsin\frac{x}{\xi}+\frac{x}{\xi}\sqrt{1-\left(\frac{x}{\xi}\right)^2}\right] \tag{4}$$

From relations 1 - 4 value of the potential barrier is determined:

$$\Delta U_2(x)=\frac{\mu_o H_c^2 l d x}{2} \tag{5}$$

for the case of $\boldsymbol{x<x_c}$, while in an opposite situation it appears relation:

$$\Delta U_3(x)=\frac{\mu_o H_c^2 l\xi^2}{2}\left[\arcsin\frac{x}{\xi}-\frac{\pi}{2}+\arcsin\frac{d}{2\xi}+\frac{x}{\xi}\sqrt{1-\left(\frac{x}{\xi}\right)^2}+\frac{d}{2\xi}\sqrt{1-\left(\frac{d}{2\xi}\right)^2}\right] \tag{6}$$

Maximum of the potential barrier is reached just in this second case. In the total energy balance additionally has been taken into account the Lorentz force arising during the transport current flow, which affect the vortices transporting magnetic flux as well as elasticity forces of the vortex array. Elasticity forces are related to the magnitude of the captured vortex deflection in the vortex lattice. Capturing vortices by the nano-sized pinning centers causes the deviation of the vortex from its equilibrium position in the regular vortex array, thus leading to an increase in the elasticity energy of the structure of the vortex lattice. This effect is the function of the deflection of the individual vortex from its equilibrium position in the lattice. We have taken it into account by assuming that an enhancement of the vortex elasticity energy is proportional to the square of the length of the vortex deflection, with coefficient of proportionality expressed by the value of the parameter $\alpha$:

$$U_{el} = \frac{2c_s \pi \xi^2 (\xi - x)^2}{l_a} = \alpha(\xi - x)^2 \tag{7}$$

Parameter $c_s$ is the corresponding elasticity shear modulus, while $l_a \approx l$ denotes the length on which the magnetic flux lines of vortices are deformed. Above model leads finally to the relation describing the potential barrier for the flux creep process $\Delta U$ in the function of the reduced current density $i = j/j_C$, where $j_C$ is defined as the critical transport current density for flux creep process. Condition for the minimum of the system energy allows therefore to deduce the final expression for the energy barrier height, which should be crossed by the vortex in the flux creep process:

$$\Delta U(i) = \frac{\mu_0 H_c^2 l \xi^2}{2} \Theta + \alpha \xi^2 \sqrt{1-i^2}\left(\sqrt{1-i^2} - 2\right) \tag{8}$$

Function $\Theta$ appearing in Eq. 8 is determined here according to the relation:

$$\Theta = \arcsin \frac{d}{2\xi} + \frac{d}{2\xi}\sqrt{1 - \left(\frac{d}{2\xi}\right)^2} - i\sqrt{1 - i^2} - \arcsin i \tag{9}$$

$H_c$ in Eq. 8 is magnetic thermodynamic critical field, $l$ pinning center thickness, while $i = j/j_c$ the reduced transport current density, as it was stated before. Parameter $\alpha$ describes the elasticity energy of the vortex lattice. $j_C$ is defined here as the transport current density $j$ for which potential barrier disappears, in the large scale limit of the pinning center size, it is for $d = 2\xi$, according to the relation:

$$j_c = \frac{\mu_0 H_c^2}{\pi \xi B} \cdot \frac{S_c (1 - S_c / a^2)}{a^2} \tag{10}$$

$S_c$ in Eq. 10 is the pinning center cross-section, while $a$ average distance between the regularly ordered pinning centers, $B$ is magnetic induction.

We insert then the expressions 8-9 into the constitutive relation, describing the generated electric field in the flux creep process, just as a function of the potential barrier height:

$$E = -B\omega a\left[\exp\left[-\frac{\Delta U_0}{k_B T}\left(1 + \frac{j}{j_C}\right)\right] - \exp\left(-\frac{\Delta U}{k_B T}\right)\right] \tag{11}$$

$\Delta U_0$ is the potential barrier height without current, $\omega$ constant value describing the characteristic flux creep process frequency, $T$ temperature and $k_B$ Boltzmann's constant, parameter $a$ is the average distance between pinning centers regularly arranged. Applying above relations we can predict the dependence of the real critical current, i.e. satisfying the electric field criterion, on the material parameters. Selected results of calculations are presented in Figs. 3-4 and indicate to the importance of the pinning centers dimensions as well as the elasticity constant $\alpha$ for the critical current magnitude.

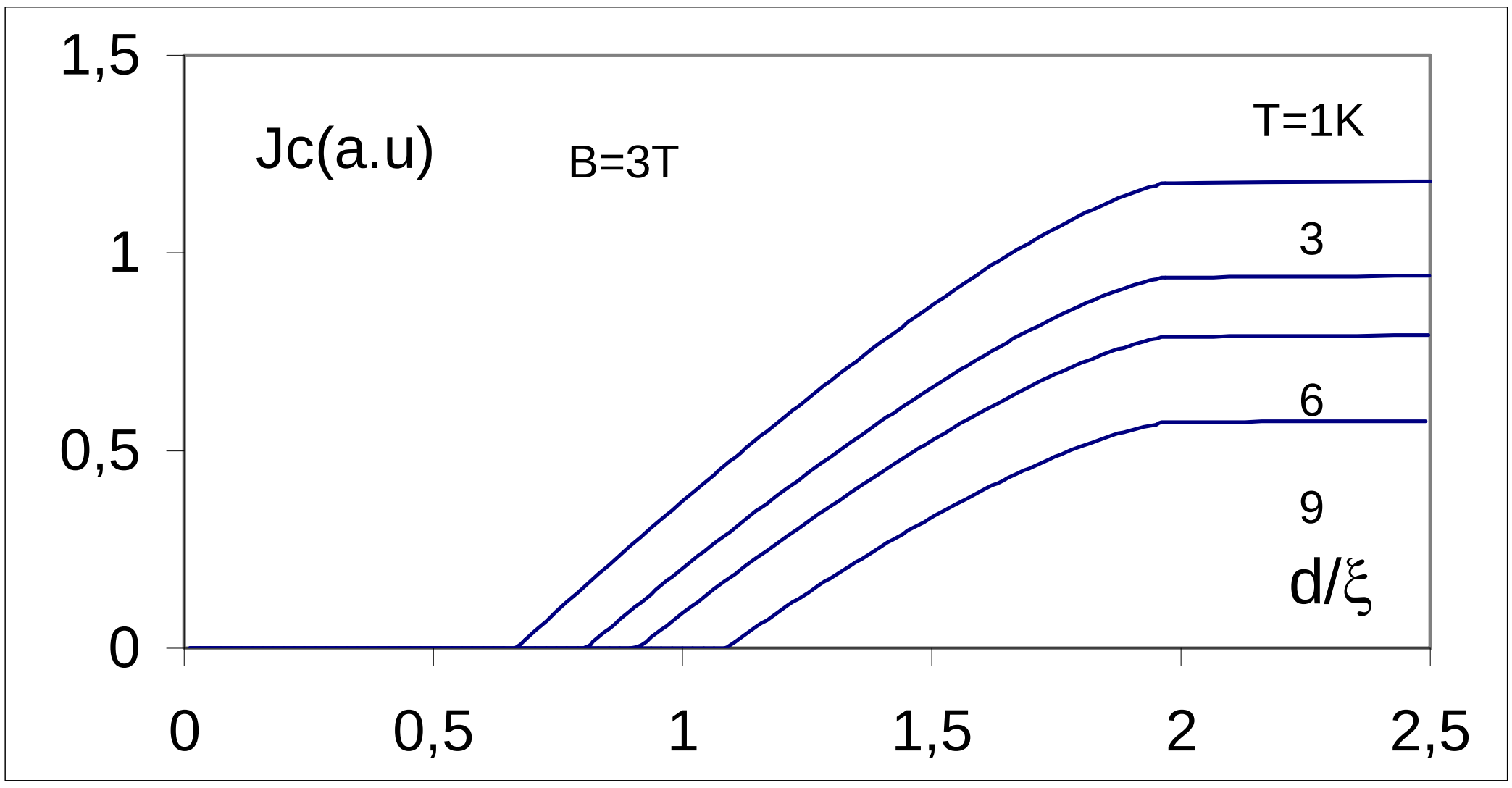

Fig. 3. The dependence of the critical current satisfying the electric field criterion versus the pinning center dimensions for various temperatures

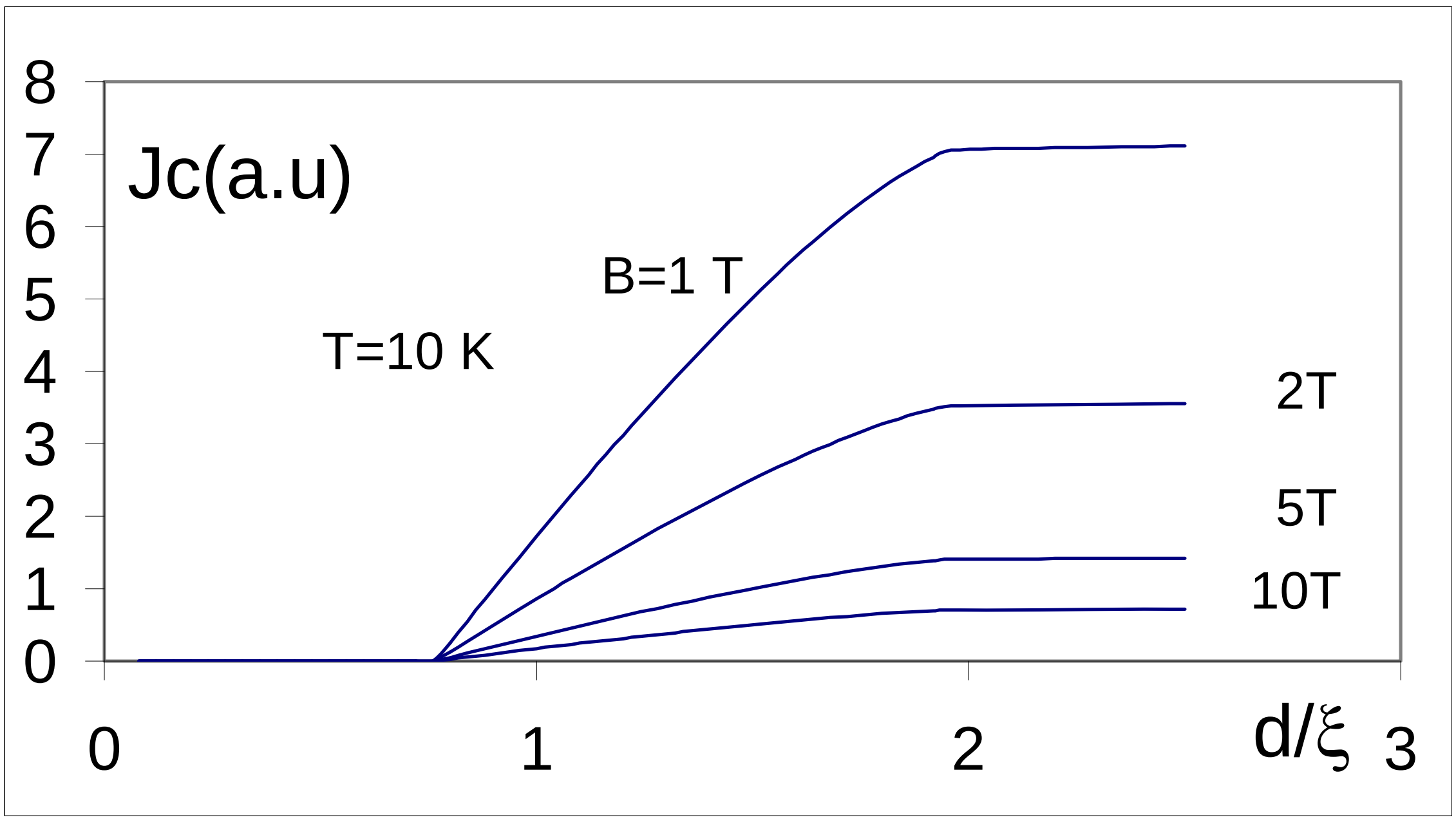

Fig.4. The influence of the nano-sized pinning centers of the width $d$, normalized to the coherence length $\xi$ on the critical current density versus applied magnetic field.

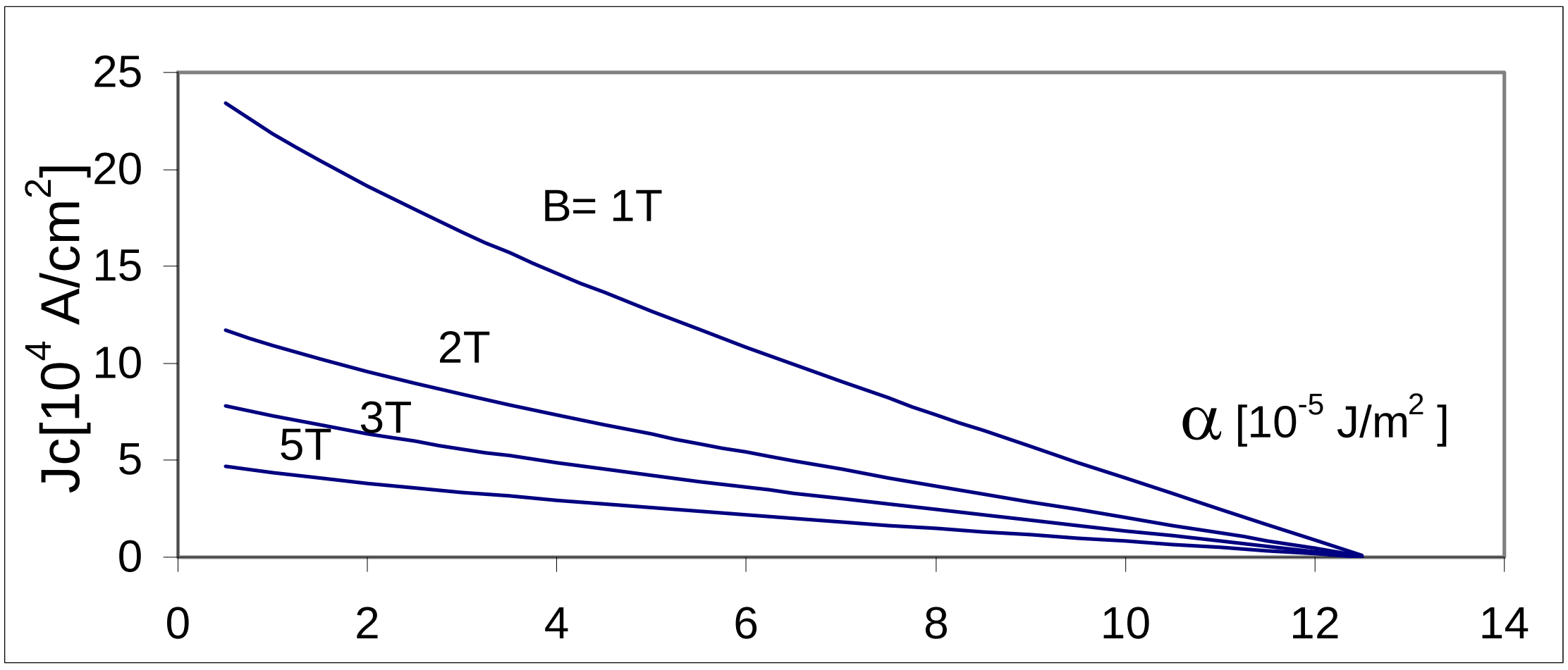

Fig. 5. The dependence of the critical current on the elasticity force of the vortex lattice (described by parameter α) versus external magnetic induction

From Figs. 3-5 we see that for low pinning centers dimensions as well as too large elasticity constant of the vortex lattice, the critical current vanishes, which result should have important meaning for technologists. In the case of the single crystalline HTc superconductors we should still remember on the layered structure of these materials. It concerns especially the Cu-based superconductors although in the recently discovered Fe-based HTc superconductors layered structure appears also. The layered crystal structure of the HTc superconductor BiSrCaCuO is presented in Fig. 6. Superconducting layers are intercalated here with the buffer ones, which supply electric charge to the $CuO_2$ layers, responsible for superconductivity effect, as shown in Fig. 5. Thickness of the layers is smaller than coherence length $\xi_o$, so due to the proximity effect the layers interact with neighboring partners or rather more precisely interact the vortices situated in the neighboring planes influencing the current-voltage characteristics. Present model allows in the first approximation to consider this interaction, which leads to the modification of the current-voltage characteristics in the individual plane, as indicates Fig. 7.

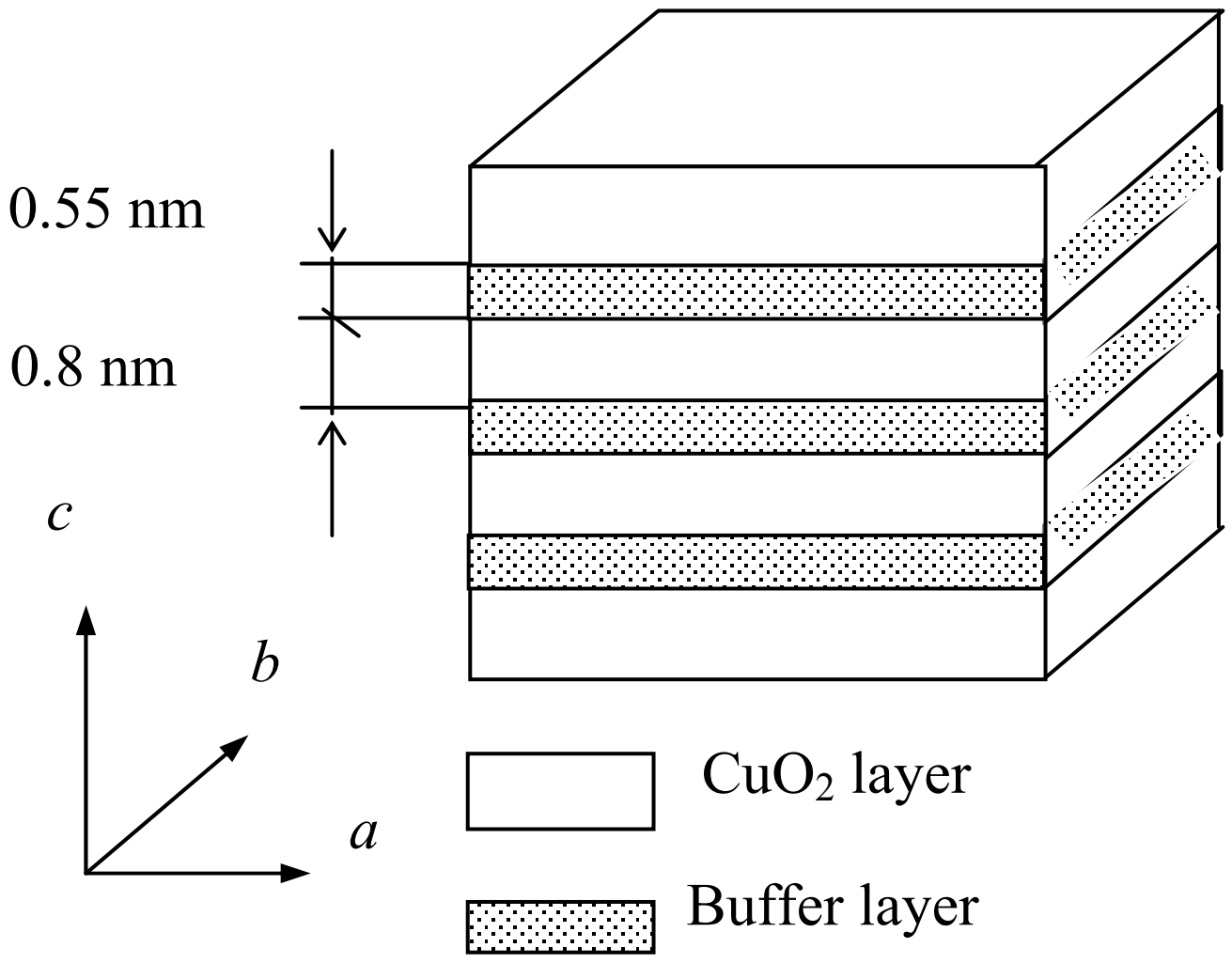

Fig. 6. Scheme of the multilayered structure of the Bi:2212 high temperature superconductor

In this Figure is shown the voltage generated under current flowing in individual plane taking into account the interacting vortices in the neighboring planes. Index ***k*** denotes just the number of interacting planes.

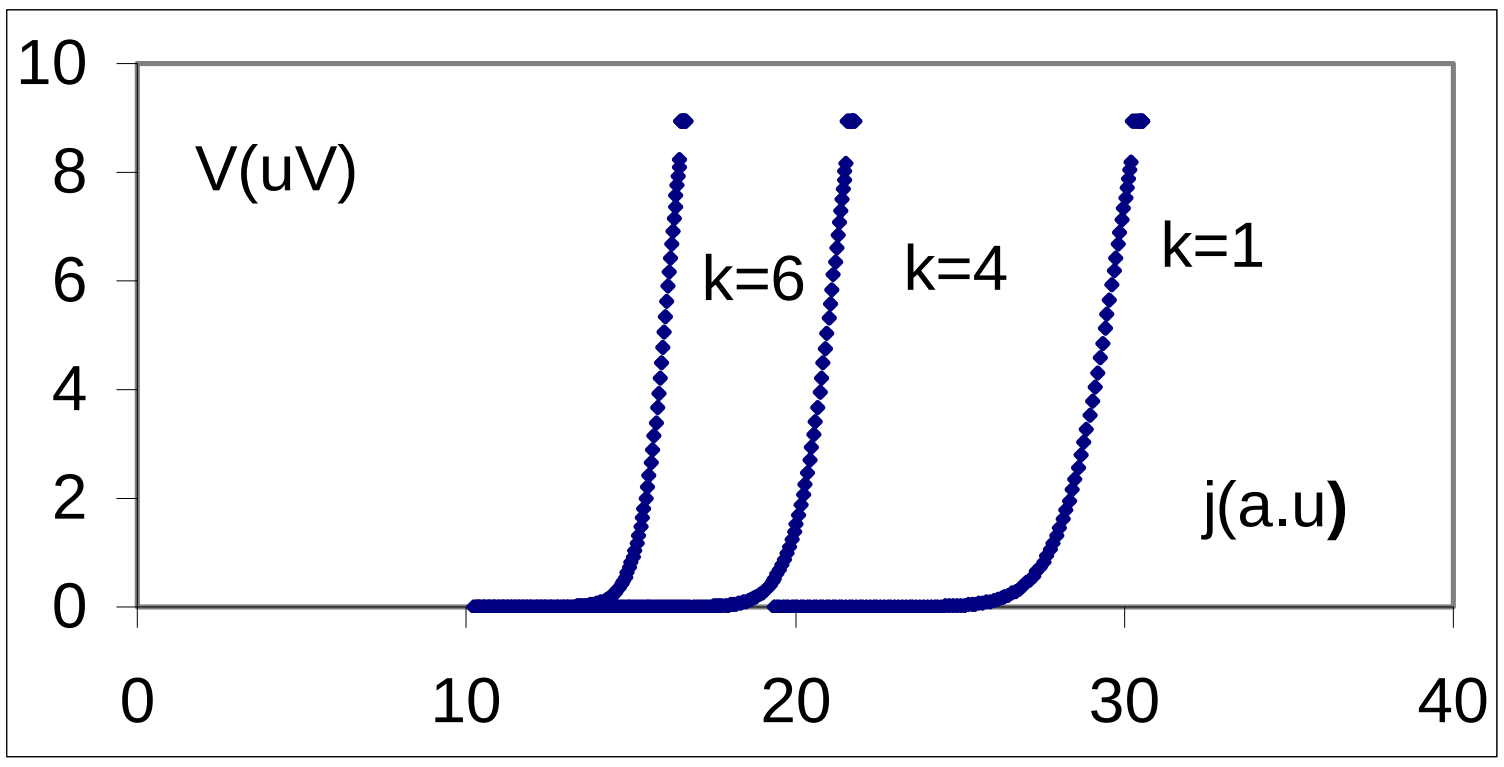

Fig. 7. The influence of the inter-plane interaction (***k*** – number of the interacting $CuO_2$ planes) on the current-voltage characteristics of the HTc layered superconductor

### 3. Comparison of the model of the pinning interaction with experimental data

Above theoretical considerations we have applied then for the comparison with an experimental data. In Fig. 8 is shown the comparison of the current-voltage characteristics calculated according to described mathematical model of the pinning interaction with the experimental data obtained for Bi-based superconducting tape. An approximation of the large pinning centers dimensions has been applied here, which means in the language of our model that it has been considered case of the $d \gg 2\xi$.

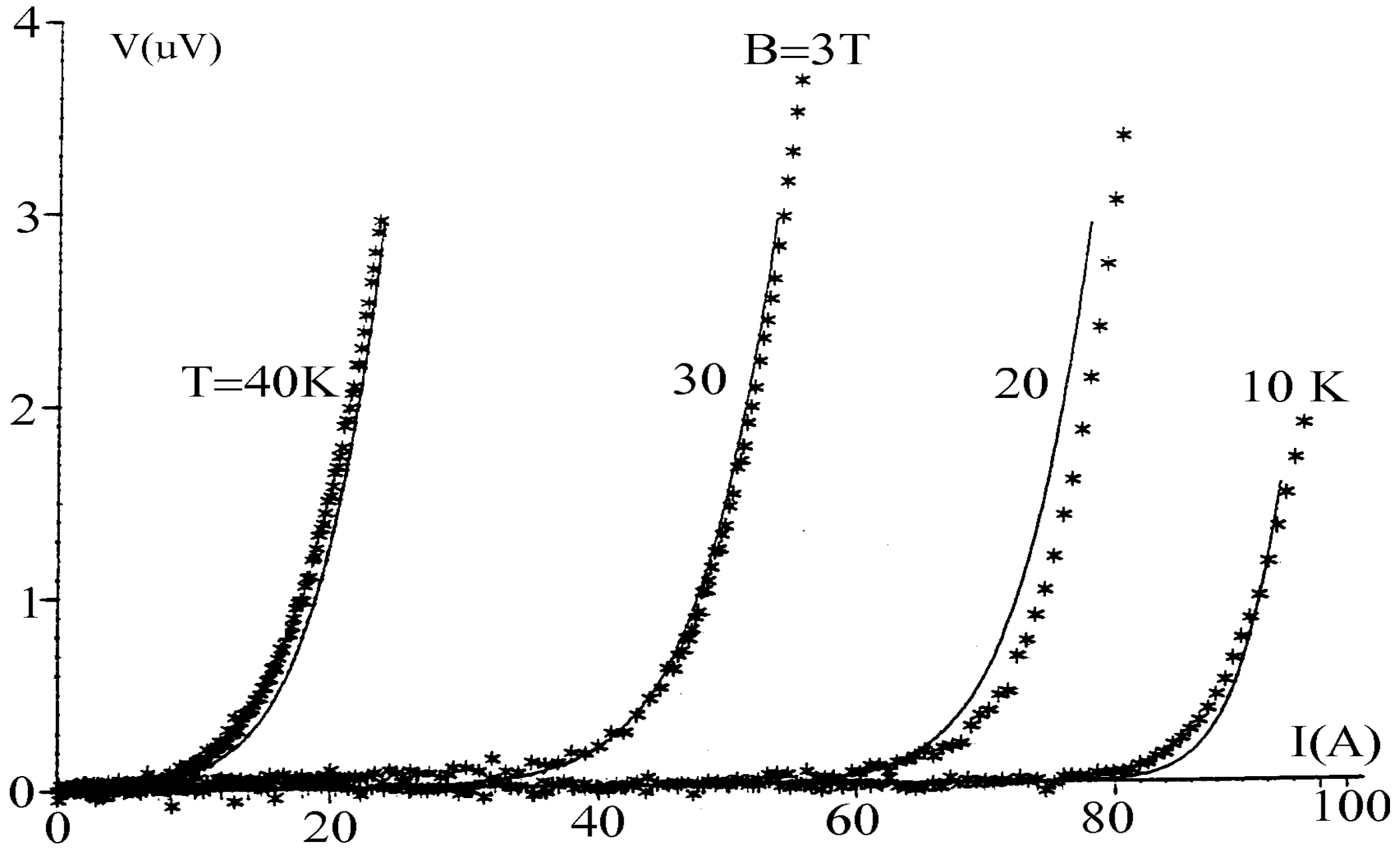

Fig. 8. Comparison of the experimental (*) and theoretical I-V curves for Bi:2223 tape

Very good agreement of theoretical model with experimental data has been received in various temperatures, while using common fitting parameter, which is the concentration of the pinning centers. It should be noticed that the defects concentration received from the scanning electron microscopy analysis indicated similar value.

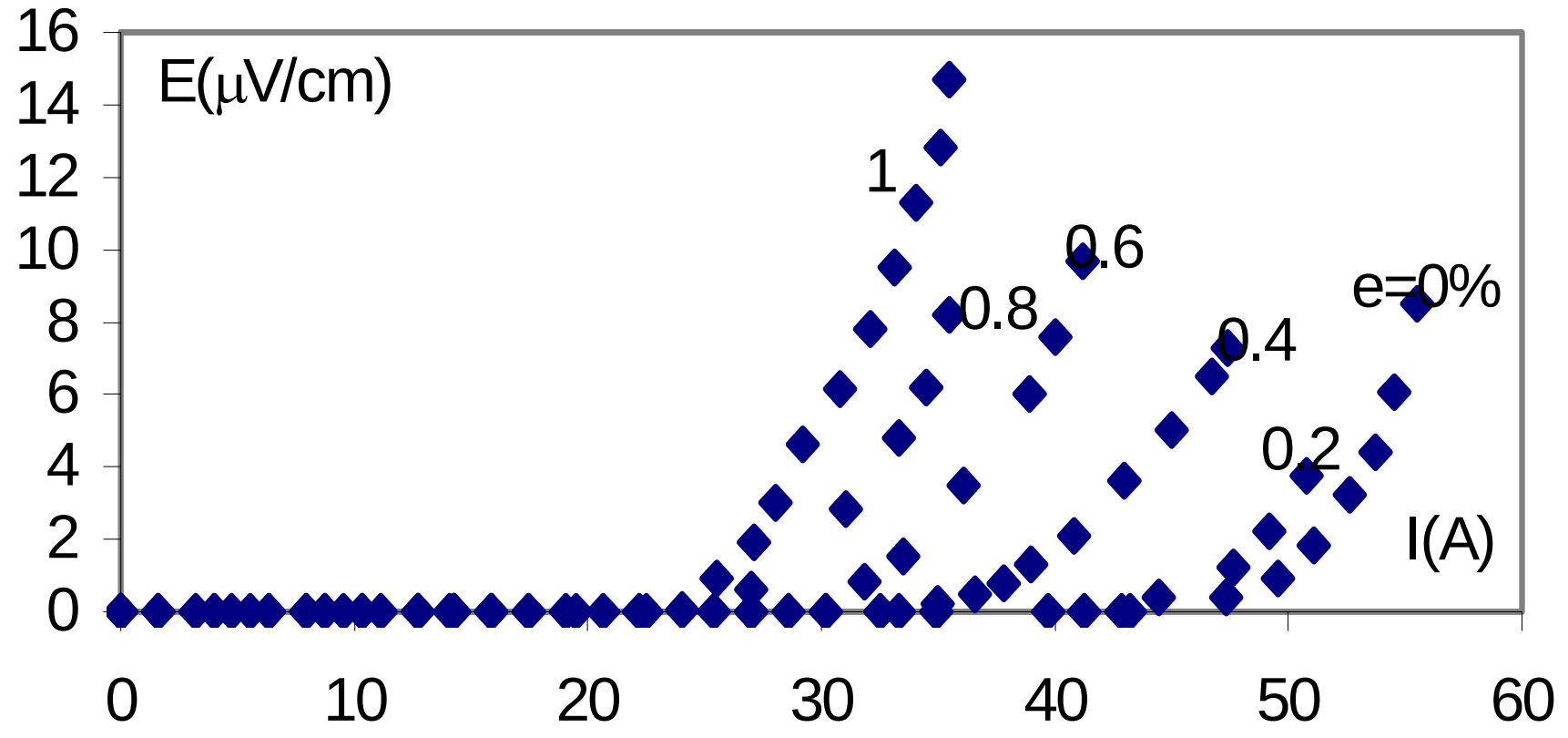

Fig. 9. Experimental dependence of the influence bending strain on the current-voltage characteristics for the superconducting Bi-2223 tape

Additionally structural defects such as micro-cracks are created as the result of the bending strain of the superconducting tapes, which effect frequently appears during winding of the superconducting coils. Experimental data of the influence bending strain on the current-voltage characteristics are given in Fig. 9, while critical current data versus bending strain shows Fig 10. Bending strain parameter ***e***, which appears in the Figs. 9-10 is defined as the ratio of the tape thickness ***t*** to its diameter ***D***: ***e= t/D***.

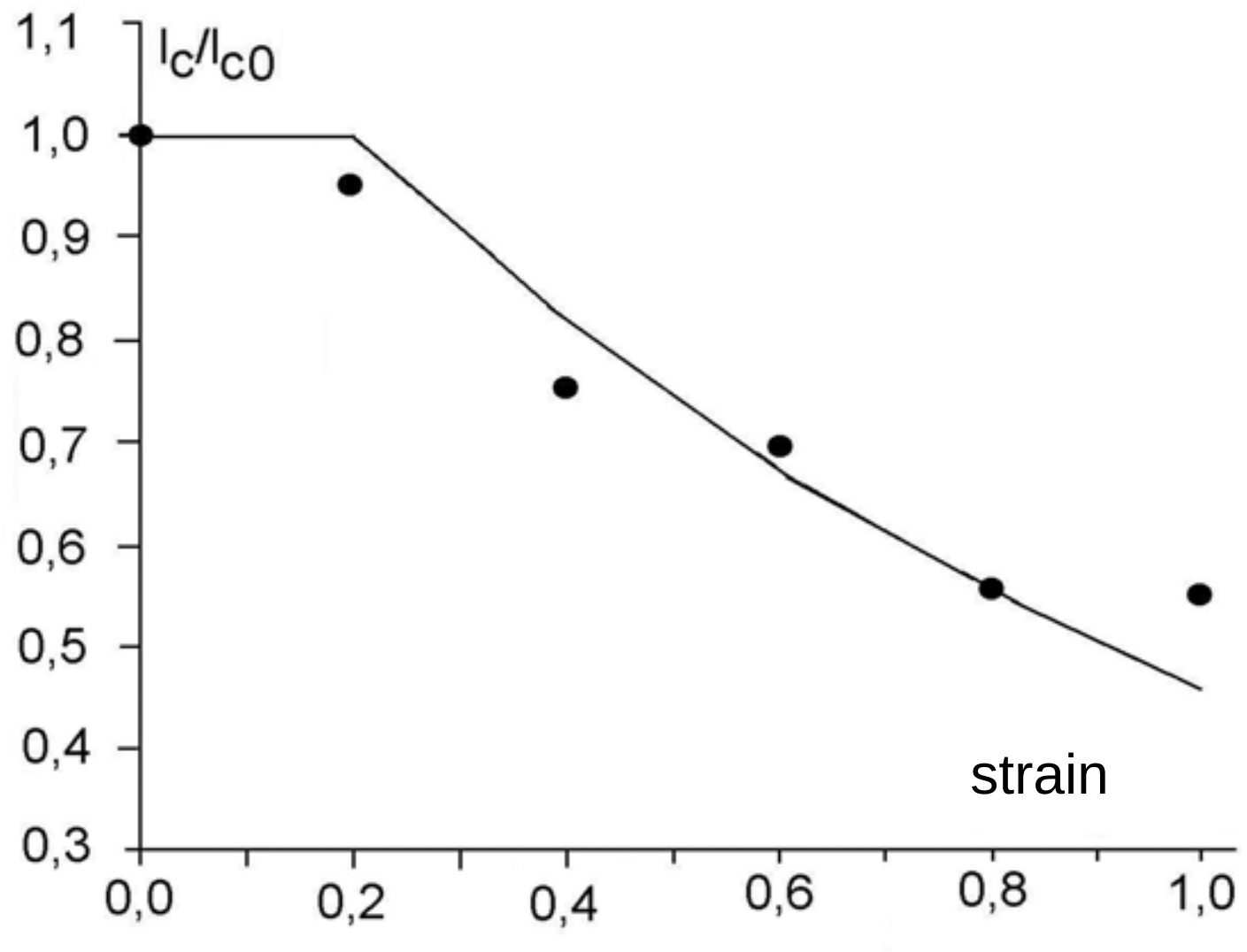

Fig. 10. Experimental dependence of the influence bending strain on the critical current of the superconducting Bi-2223 tape

In Fig. 11 are presented calculated according to the present model the theoretical influence of the bending strain on the current-voltage characteristics. These theoretical results are in qualitative agreement with the experimental data shown in Fig. 9.

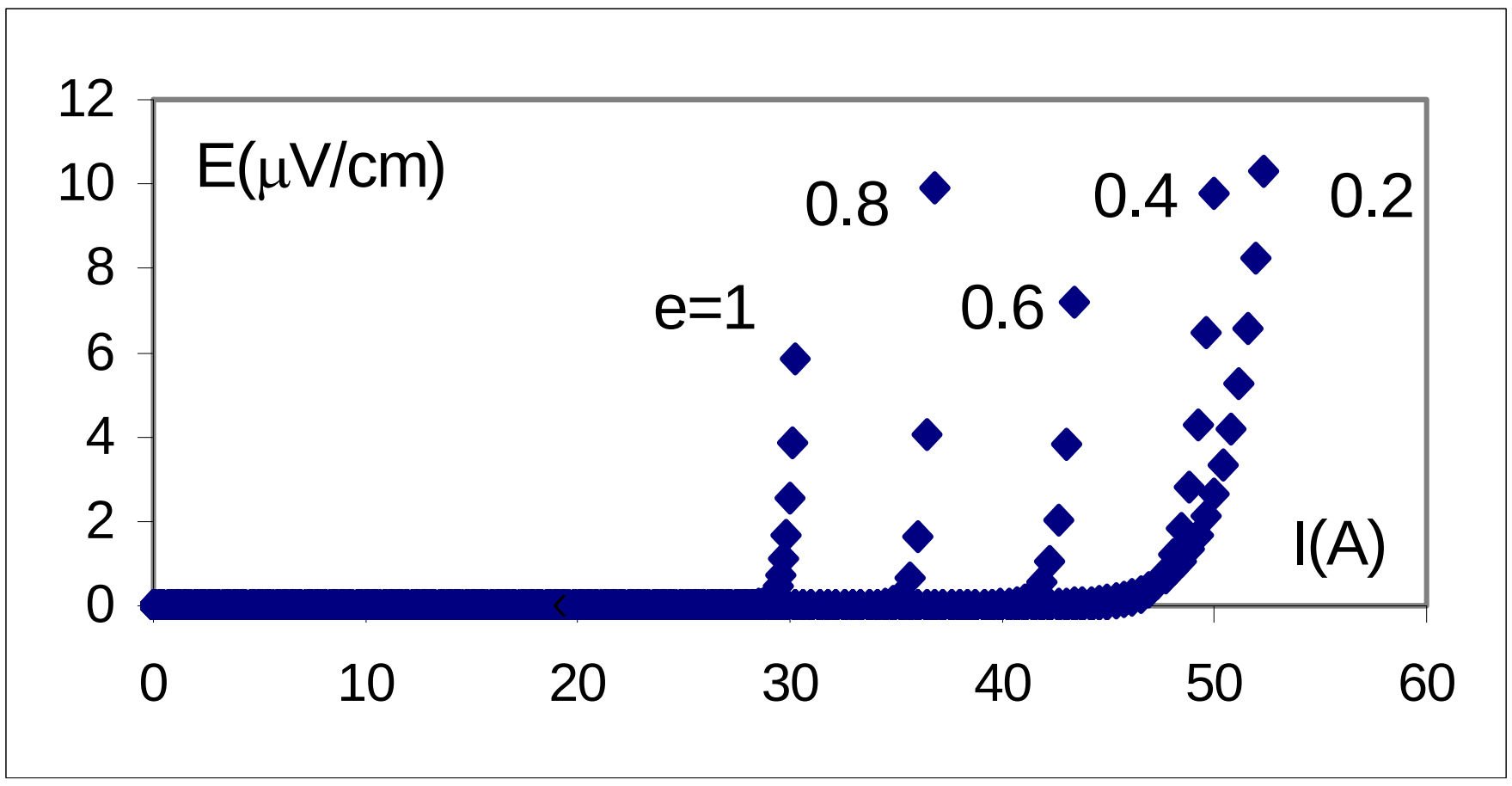

Fig. 11. Theoretically predicted influence of the bending strain ***e*** (in percents), on the current-voltage characteristics for Bi:2223 tape

Knowledge of the critical current allows us to determine the volume pinning force, it is force acting on vortices in the unit volume. To determine this force the many body problem of the occupation varying with magnetic field number of the vortices on the fixed number of the pinning centers should be taken into account. Final result of the theoretical magnetic field dependence of the pinning force for three temperatures is shown in Fig. 12. In Fig. 13 from other side it is shown experimental dependence of the pinning force versus magnetic field, measured for Bi-2223 tape. Also in this case good agreement of the model with experiment has been received, which confirms theoretical proposal. The shift of the maximal value of the pinning force with temperature observed experimentally, as indicates Fig. 13 has the place too in the theoretical model, which is shown in the Fig. 12. Presented model allows to explain also the dynamic anomalies of the current-voltage characteristics of the HTc ceramics observed us previously [10].

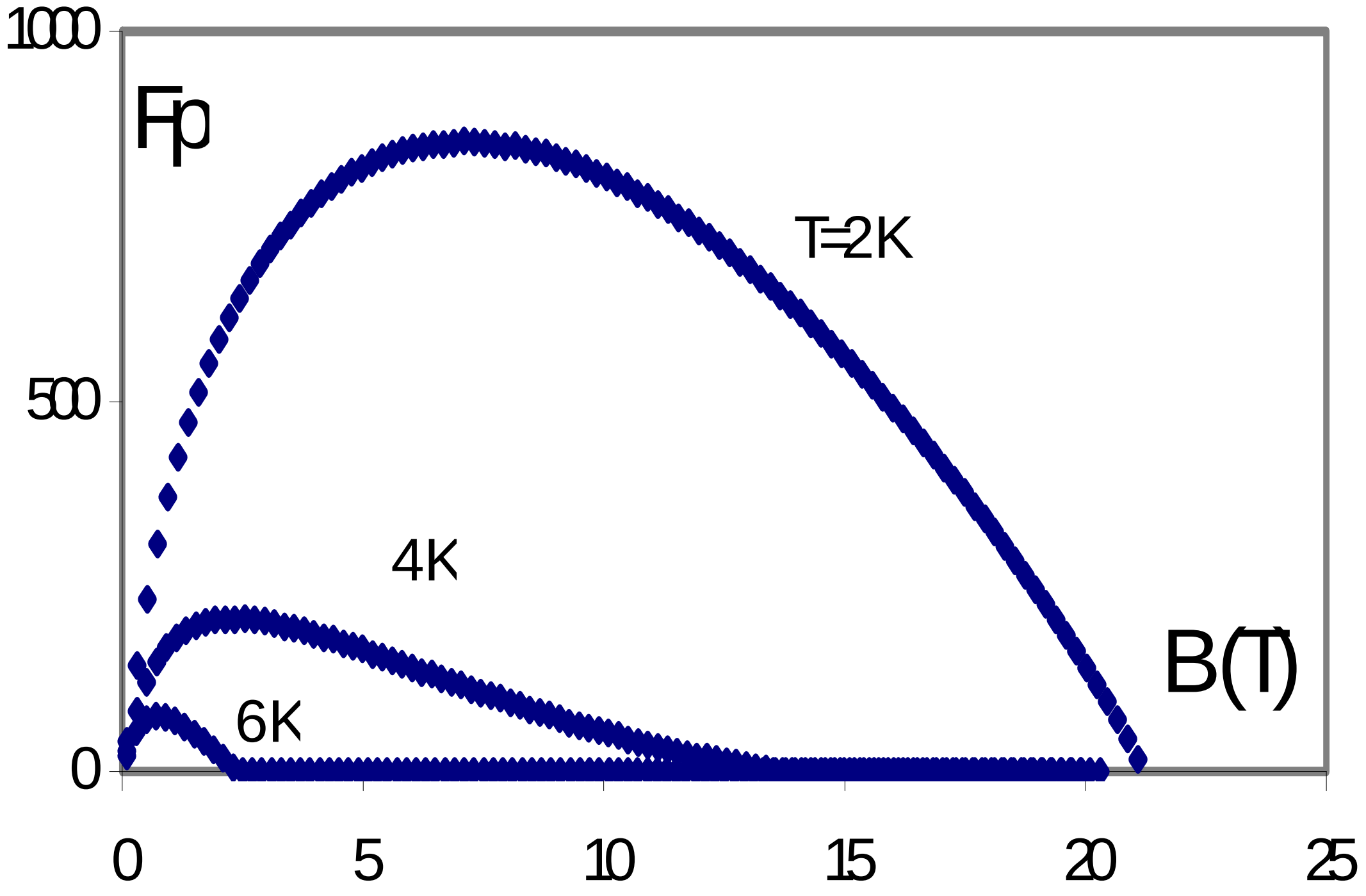

Fig. 12. Theoretical dependence of the flux pinning on the magnetic induction versus temperature

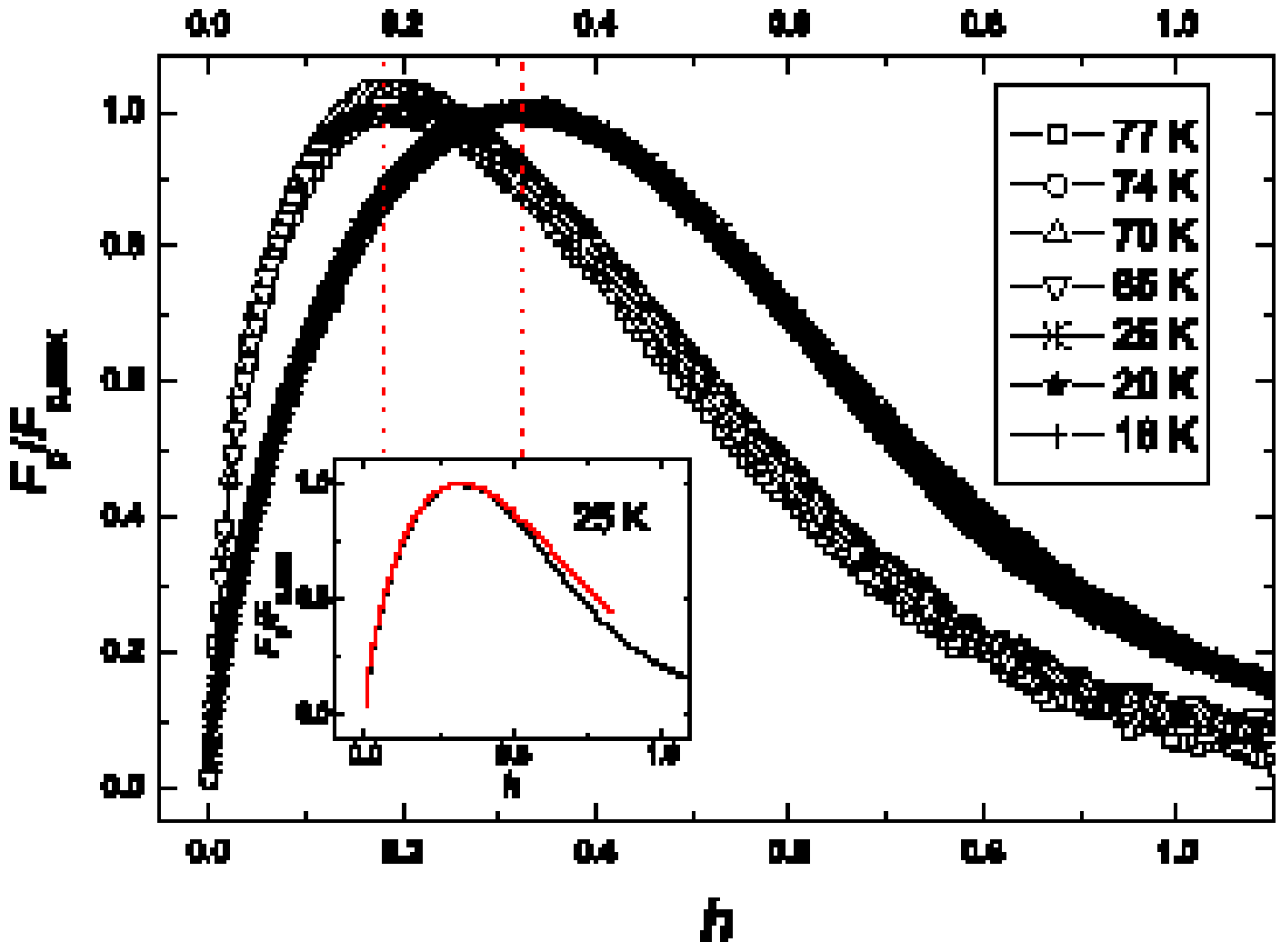

Fig. 13. Experimental dependence of the normalized flux pinning force $F_p$ for Bi-2223 tape on magnetic induction, for various temperatures [10]

## 4. Influence of the initial captured vortex position on the critical current density

In detail analysis of the pinning interaction we should still to consider the mechanism of vortex capturing on the pinning centers. Two topics are therefore regarded now, connected with the geometrical aspects of this capturing process. First one is related to the initial position of the captured vortex, while second to the renormalization of the screening currents around the vortex captured on the nano-sized pinning center. Now we consider this first point, while second one will be the subject of the next clause 5. In the previous point it has been considered the vortex dynamics problem for the initial state of the vortex captured onto the depth of the renormalized coherence length, as shows Fig. 2. Such configuration is preferred because it should allow among other still for the flow of the superconducting screening currents, around the vortex core, due to the proximity effect and keep this way the vortex structure. For evaluating the relevance of this initial configuration for the final results describing the potential barrier height and critical current density, the independent calculations have been performed for other terminal initial vortex state. We have considered therefore additionally the configuration of the fully captured vortex, it is vortex core pinned on the depth of 2ξ inside the pinning center of the nanometric size. The basic equations describing the initial state normal energy for the fully captured vortex as well as for vortex shifted against this initial position have been derived according to the method presented in the point 2 while results are given below. Initial normal state energy of the fully captured vortex, according to the notation of the Fig. 2 is equal:

$$U(-\xi) = \frac{\mu_0 H_c^2 l_p}{2}\left[\pi\xi^2 - 2\xi^2 \arcsin\frac{d}{2\xi} - d\sqrt{\xi^2 - \frac{d^2}{4}}\right] \tag{12}$$

For the first deflection of the vortex from this initial position it appears the dependence:

$$U_1(x) = \frac{\mu_0 H_c^2 l_p}{2} \left[ \begin{array}{l} \pi\xi^2 - 2\xi^2 \arcsin\frac{d}{2\xi} + \\ \xi^2 \arcsin\frac{\sqrt{\xi^2 - x^2}}{\xi} - d\sqrt{\xi^2 - \frac{d^2}{4}} \\ + x\sqrt{\xi^2 - x^2} \end{array} \right] \tag{13}$$

valid for the following range of the vortex deflection:

$$-\xi \le x \le -\xi\sqrt{1 - \left(\frac{d}{2\xi}\right)^2} \tag{14}$$

For the next range of the vortex movement described by the new boundary condition:

$$-\xi\sqrt{1-\left(\frac{d}{2\xi}\right)^2} \le x \le \xi\sqrt{1-\left(\frac{d}{2\xi}\right)^2} \tag{15}$$

the normal state energy of the shifted vortex is given by the relation:

$$U_2(x) = \frac{\mu_0 H_c^2 l_p \xi^2}{2}\left[\pi - \arcsin\frac{d}{2\xi} - \frac{d}{2\xi}\sqrt{1-\left(\frac{d}{2\xi}\right)^2} + dx\right] \tag{16}$$

For still larger vortex shift normal state potential is:

$$U_3(x) = \frac{\mu_0 H_c^2 l_p \xi^2}{2}\left[\frac{\pi}{2} + \arcsin\frac{x}{\xi} + \frac{x}{\xi}\sqrt{1-\left(\frac{x}{\xi}\right)^2}\right] \tag{17}$$

Expressions for the potential barrier height, which determine the current-voltage characteristics have been derived then and are given below for each of the considered cases:

$$\Delta U_1(x) = \frac{\mu_0 H_c^2 l_p}{2}\left[\xi^2 \arcsin\frac{\sqrt{\xi^2 - x^2}}{\xi} + x\sqrt{\xi^2 - x^2}\right] \tag{18}$$

for the values of the vortex shift described by parameter x, in the range given by Eq. 14.

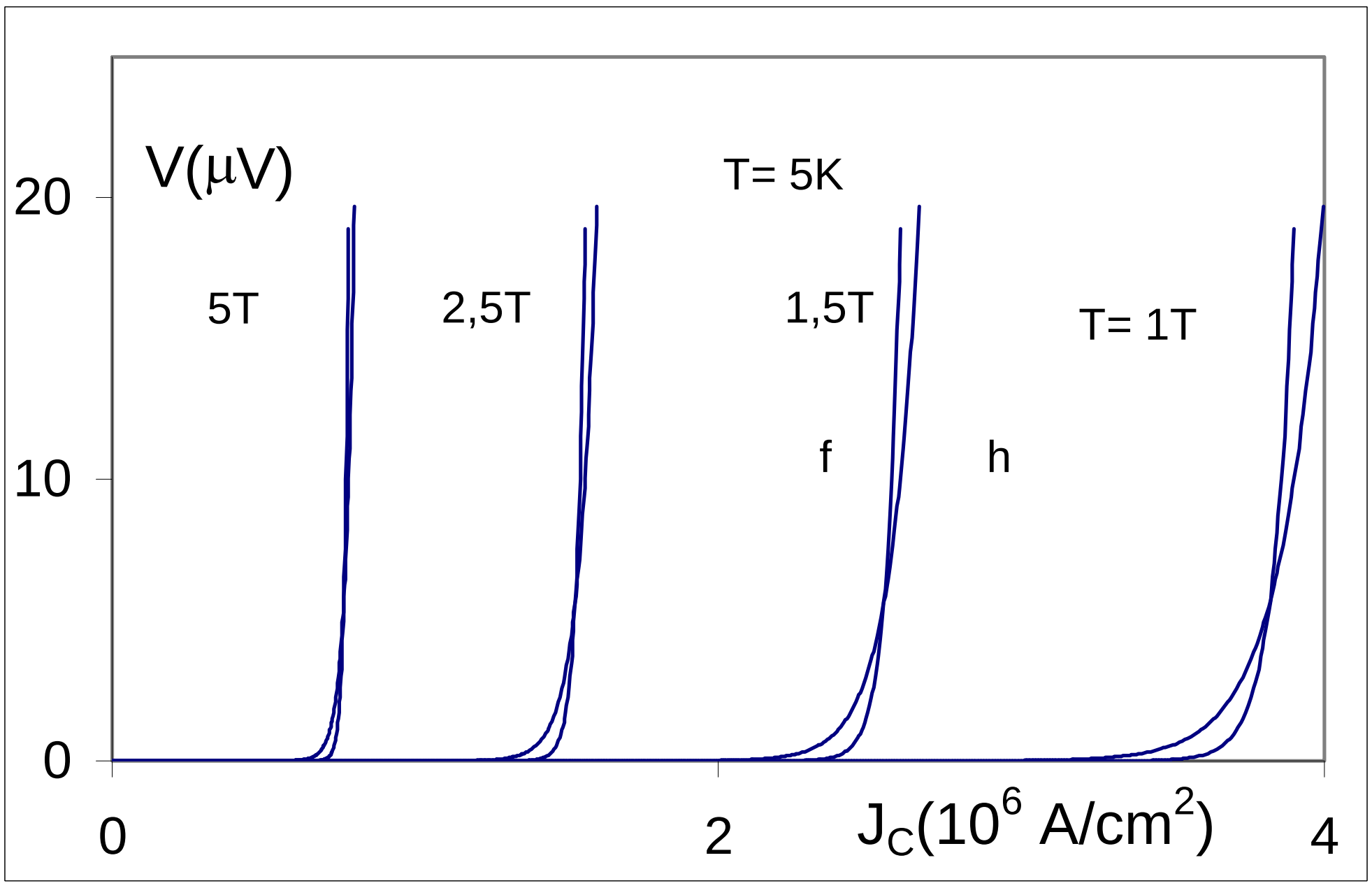

Fig. 14. The current-voltage characteristics of the HTc superconductor with half pinned vortex (h) in initial state and fully pinned (f) for d/2ξ=0,4 versus applied external magnetic field.

For larger shift of the vortex in the range defined by Eq. 15, potential barrier height is equal:

$$\Delta U_2(x) = \frac{\mu_0 H_c^2 l_p}{2}\left[\xi^2 \arcsin\frac{d}{2\xi} + \frac{d}{2\xi}\sqrt{\xi^2 - \frac{d^2}{4}} + dx\right] \tag{19}$$

For still higher vortex deflection potential barrier height is described by the relation:

$$\Delta U_3(x) = \frac{\mu_0 H_c^2 l_p}{2}\begin{bmatrix} -\frac{\pi\xi^2}{2} + \xi^2 \arcsin\frac{x}{\xi} + \\ x\xi\sqrt{1 - \left(\frac{x}{\xi}\right)^2} + 2\xi^2 \arcsin\frac{d}{2\xi} \\ + d\sqrt{\xi^2 - \left(\frac{d}{2}\right)^2} \end{bmatrix} \tag{20}$$

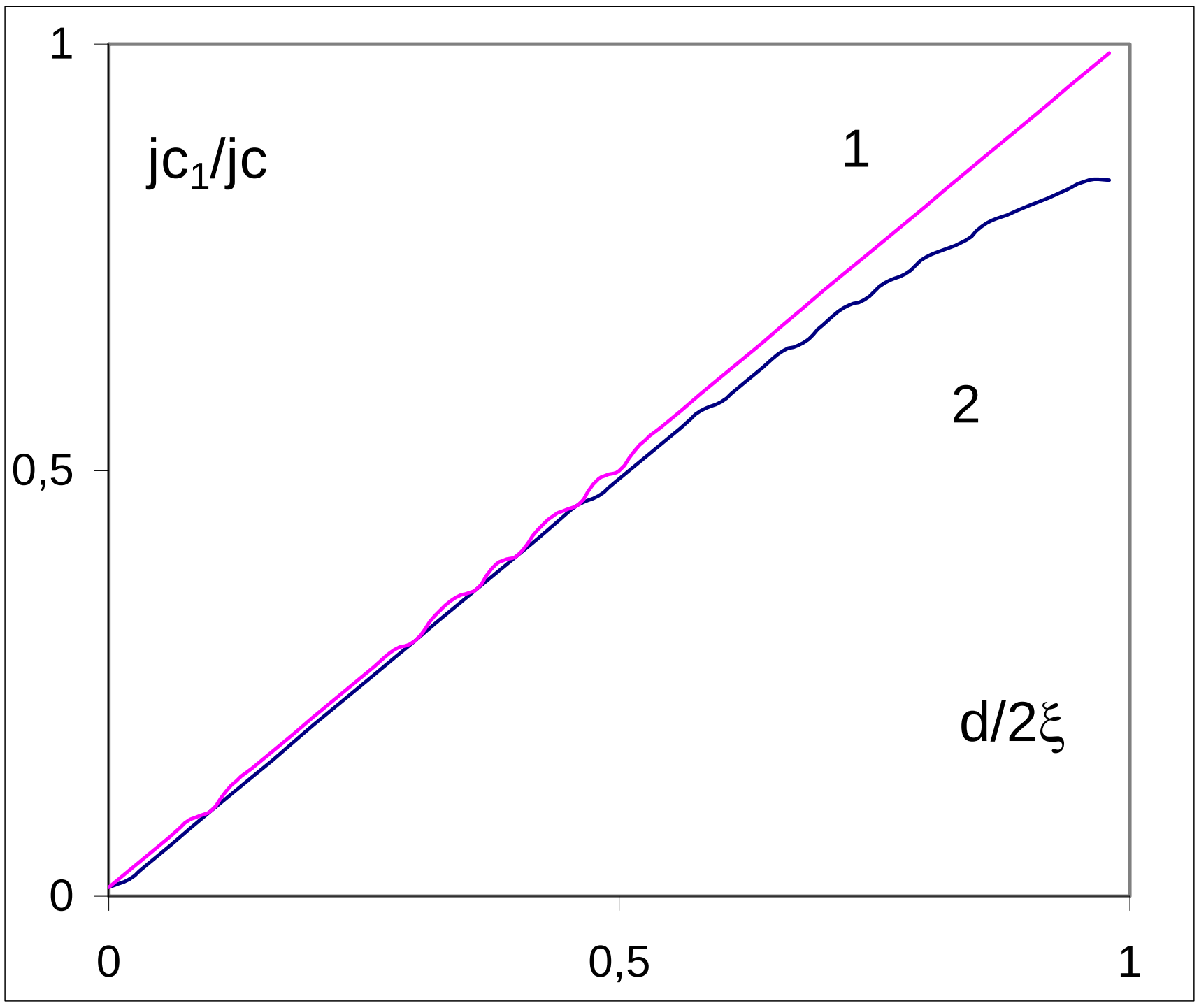

Fig. 15. The dependence of the reduced critical current density on the pinning center width, for two initial positions of the vortex: 1 – half captured, 2 – fully captured.

Transforming above dependences 18 - 20 from translational into the current representation, in the case of the full vortex capturing we obtain final relation describing the potential barrier height in the function of the current density $\boldsymbol{i = j/j_c}$, reduced to the critical one:

$$\Delta U_3(i) = \frac{\mu_0 H_c^2 l_p \xi^2}{2} \begin{bmatrix} -\arcsin(i) - i(2+\sqrt{1-i^2}) + \\ 2\arcsin\frac{d}{2\xi} + \frac{d}{\xi}\sqrt{1-\left(\frac{d}{2\xi}\right)^2} \end{bmatrix} \tag{21}$$

Eq. 21 we insert then into the constitutive equation 9 describing the current-voltage characteristics in the function of the potential barrier height $\boldsymbol{\Delta U}$. The results of the comparison of the calculations of the current-voltage characteristics for both (fully captured and half captured) initial states are shown in the Fig. 14. For better recognition of the behavior of the considered model describing current-voltage characteristics of HTc superconductors, the calculations have been performed as previously for various magnetic fields. As follows from Fig. 14 for both cases it is of the fully captured vortex as well as for half-captured the difference in the current - voltage characteristics are at least in this case negligible.

Influence of an initial position of the captured vortex on the critical current of the HTc superconductor in reduced units versus pinning center width is presented in Fig. 15. The critical current density $j_{c1}$ shown in this figure is defined as the current density leading to the disappearance of the energy barrier $\Delta U$ in that case, while $j_c$ has been defined previously in Eq. 5.

Fig. 16 from other side shows the comparison of the reduced potential barrier versus the current normalized to the critical one for various pinning forces models [12-13]. The results obtained in the present approach are marked by numbers 3-5 here and are in fact similar in both limiting cases of the full and half captured vortices in the initial state.

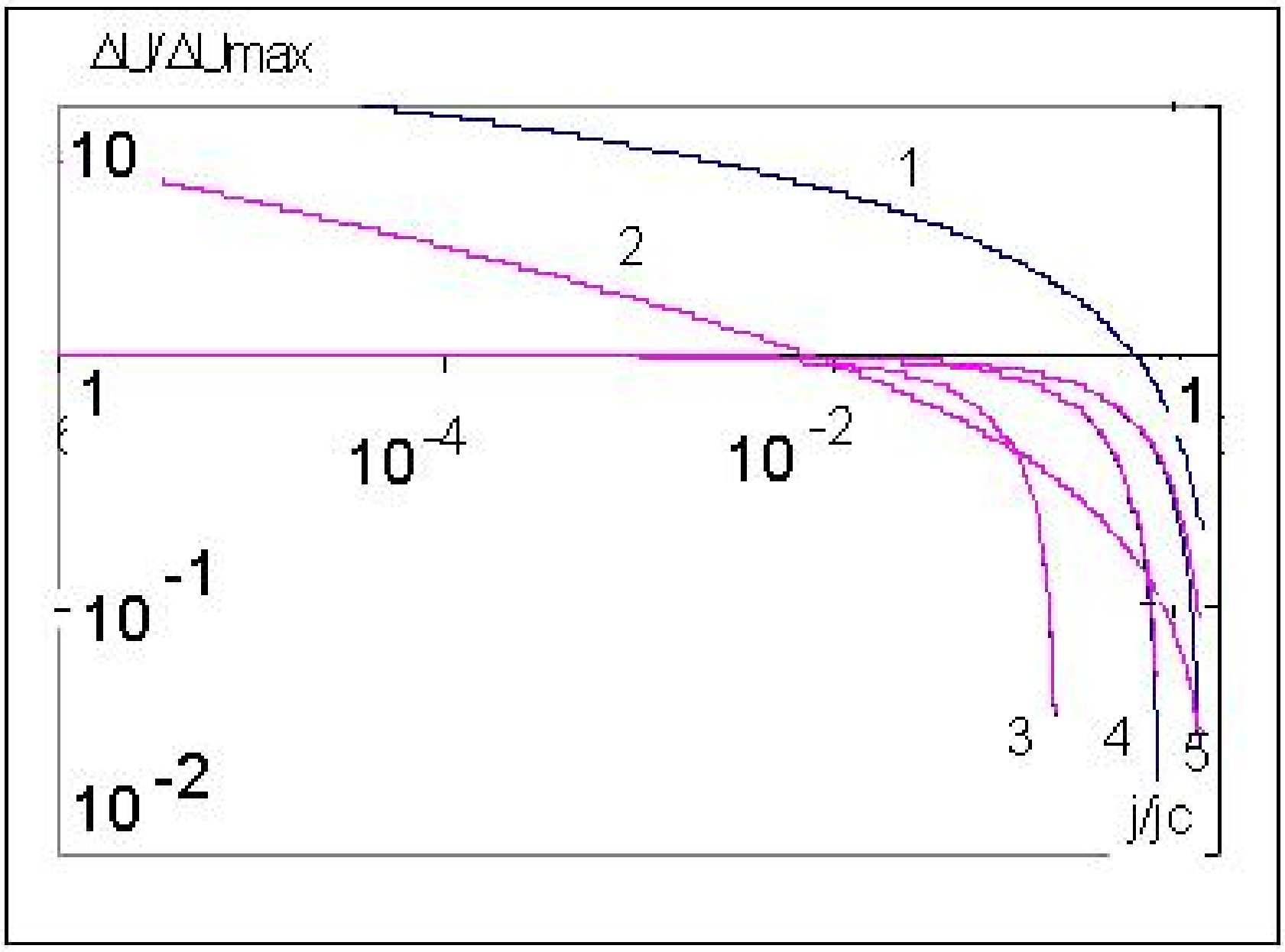

Fig. 16. The dependence of the reduced potential barrier on the current normalized to the critical one for: (1) logarithmic dependence ln (j/jc) of the potential barrier, (2) dependence of the form $(j/jc)^{-1/7}$-1, (3) according to the present model for d/2ξ = 0,15, (3) according to the present model for d/2ξ = 0,5 (3) according to the present model for d/2ξ = 0,95

**5. Influence on the critical current of the modification of the screening current distribution in the captured vortex**

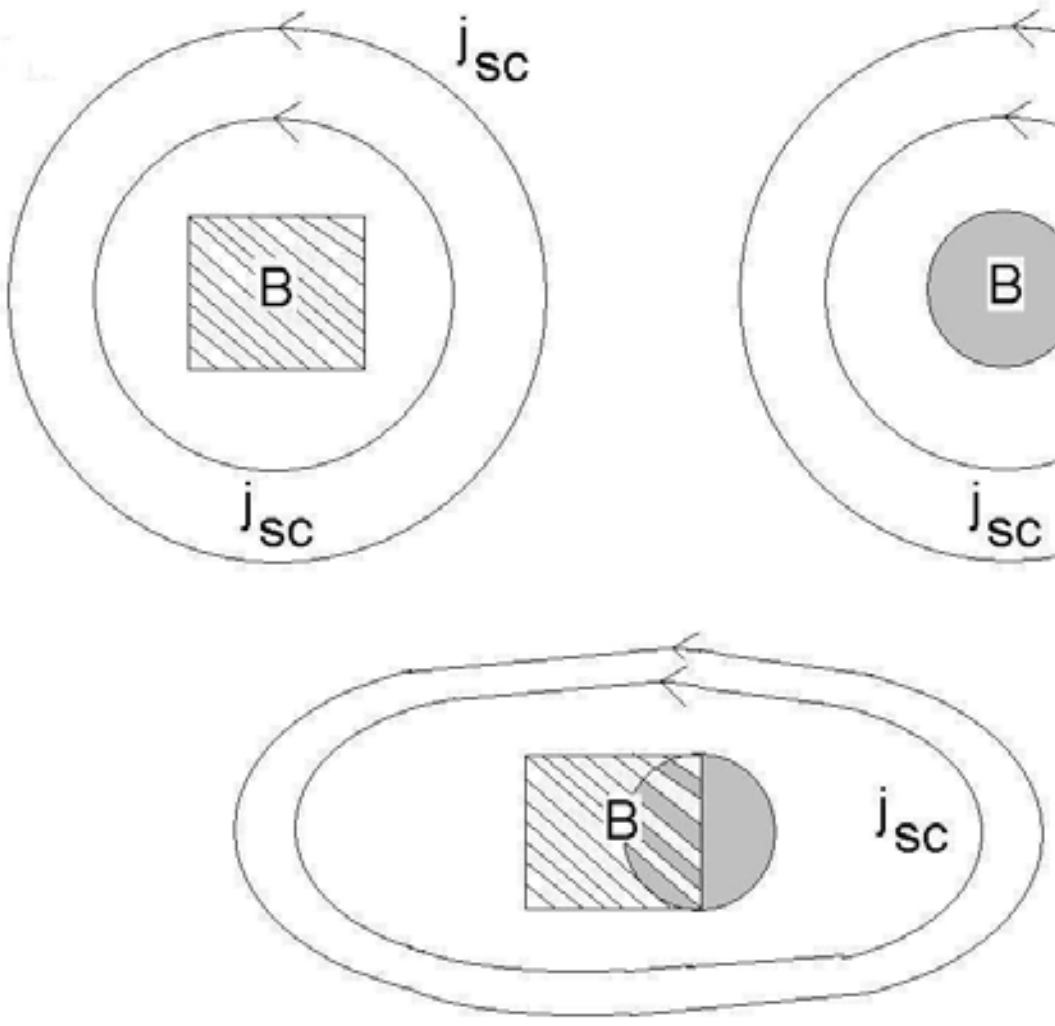

Fig. 17. Schematic comparison of the screening currents distribution and magnetic induction penetration into the vortex core and isolated nano-sized defect (up) and for the captured vortex (down) in HTc superconductor

In previous points we have considered the model describing the pinning interaction based on the energy balance, in which has been applied an assumption of an unchanged electromagnetic shape of an isolated, free vortex and for the vortex captured on the nano-sized pinning center. In fact however we should remember that the vortex is composed from the circulating currents around the vortex core. In captured vortex the screening current distribution is modified as indicates schematically Fig. 16.

Generally description of such configuration is very complicated from mathematical point of view subject. In the present paper we have applied therefore very simplified approach, in which has been considered according to Fig. 17 the generalized magnetic flux quantization. For an isolated vortex and nano-sized pinning center the flux quantization conditions are filled separately for both objects. However for the captured vortex the screening currents distribution is renormalized, while quantization condition concerns now the total object of the captured vortex and pinning center. This very complicated effect in the first approximation we have described by an assumption of the modification of the induction distribution in the vortex core of the captured vortex. We have considered therefore the case in which magnetic field penetrates the superconducting material not only through the magnetic vortices but also by the normal state inclusions, observed just in the scanning electron microscopy, for instance holes and normal grains. Above geometry presents Figure 18, which shows the pancake vortex captured on the nano-pinning center of rectangular shape. If vortex leaves the pinning center then the flux inside it is quantized and brings the value of $\Phi_0 = 2{,}067 \cdot 10^{-15}$ Wb. For captured vortex magnetic flux penetrating its core should be treated as the superposition of the magnetic field passing the normal state inclusion, which is proportional to external applied magnetic field and quantized vortex magnetic flux. It is generally very complicated effect, connected with the variation of the current distribution, because screening currents now surround both the vortex core as well as nano-sized defect. We notice too that there appear sometimes the suggestions of transporting by the vortex not only individual flux quantum but its multiplication, despite of non - favour energetically such configuration, which ideas are partly similar to presented in this model.

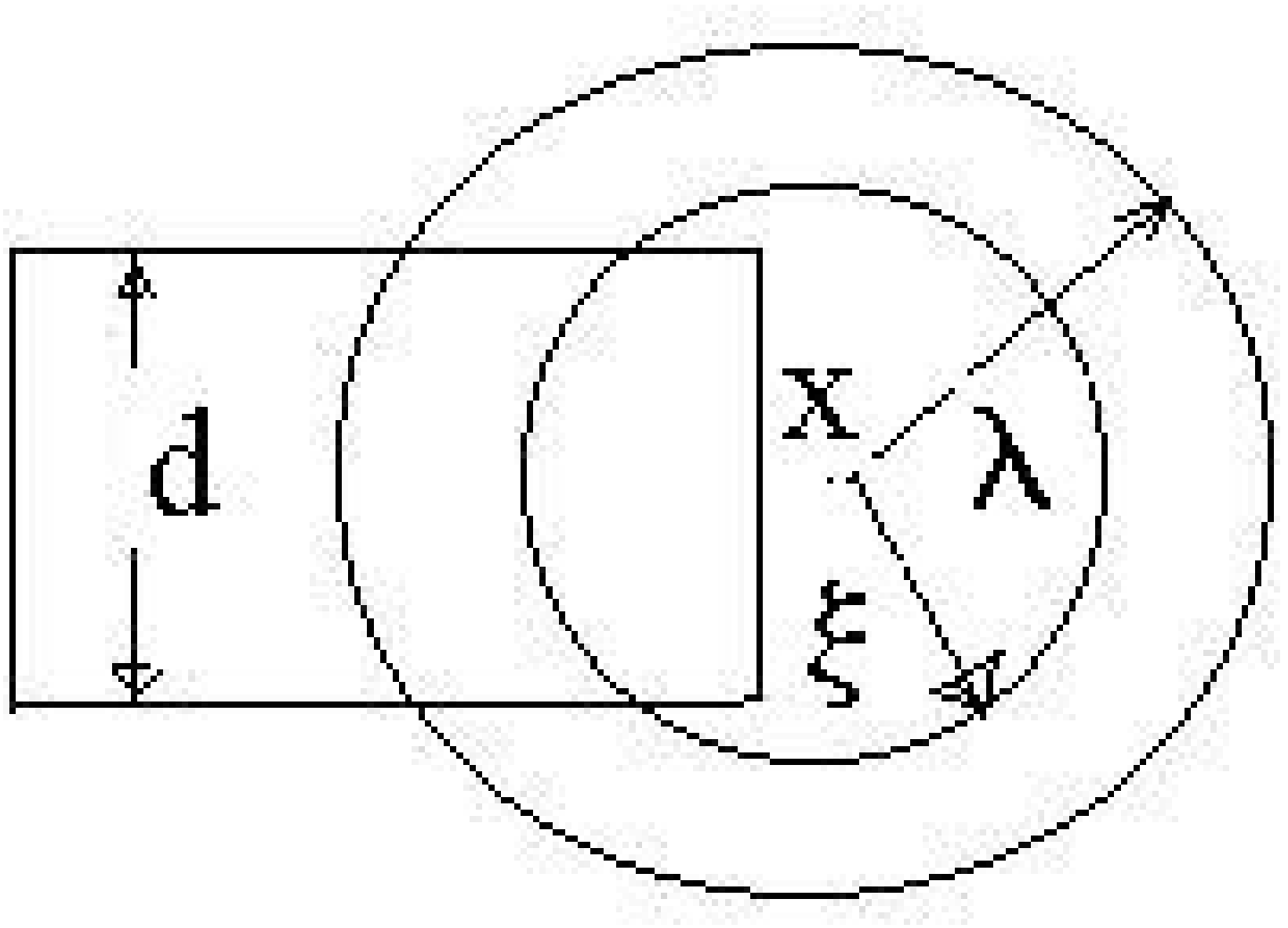

Fig. 18. View of the vortex captured on the nano-sized pinning center, of the core radius equal to coherence length $\xi$ and full dimension of the size of the penetration depth $\lambda$.

In this first approach we apply here rather rough approximation of the unchanged cylindrical symmetry of the vortex structure even in the captured state. If we approximate further the magnetic induction profile in the vortex, through the London's relation:

$$B(r) = B(0)\exp(-\frac{r}{\lambda}) \tag{22}$$

then the flux quantization condition leads to the following integral:

$$\Phi_0 = 2\pi\int_0^\infty B(0)\exp(-\frac{r}{\lambda})rdr = 2\pi\lambda^2 B(0) \tag{23}$$

***B(0)*** is magnetic induction in the middle of the vortex core, which now is simply proportional to the total magnetic flux in the vortex:

$$B(0) = \frac{\Phi_0}{2\pi\lambda^2} \tag{24}$$

Eq. 24 indicates clearly that in above first approximation the magnetic flux carried by the vortex determines the magnetic induction in the center of vortex. It is especially important for the pinned vortex, which geometry is presented in the Fig. 18. Then the flux passing the captured vortex area, is according to the present, simplified model significantly modified in the respect to the conditions 23-24. Magnetic flux equal to $\Phi_0$ is quantized in isolated vortex, while in captured one, due to the significant modification of the electromagnetic circuit quantized should be already total flux passing through the vortex core and magnetic flux crossing the pinning center. In the first approximation we have assumed therefore, according to these assumption and scheme shown in Fig. 18, that the magnetic flux in the region of the captured vortex is enhanced in the comparison to an isolated one. The magnetic flux in the captured vortex, is then superposition of the flux quantum and of magnetic field appearing in the common region of pinned vortex and empty hole or normal metal inclusion, acting as the pinning center. As it was stated previously, we assume further additionally that the cylindrical geometry of the pinned vortex retains unchanged. In the initial state of the pancake vortex captured on the pinning center, as it shows Fig. 18 the enhanced magnetic flux $\Phi_1$ in the vortex area is therefore according to these considerations equal:

$$\frac{\Phi_1}{\Phi_0} = 1 + \frac{B_e\lambda^2}{\Phi_0}\left[\arcsin\frac{d}{2\lambda} + \frac{d}{2\lambda}\sqrt{1-\left(\frac{d}{2\lambda}\right)^2}\right] \tag{25}$$

where $\boldsymbol{B_e}$ is magnetic induction, passing through the pinning center, while ***d*** denotes the width of the pinning center of the rectangular shape. Parameter $\boldsymbol{B_e}$ is roughly proportional to an applied external magnetic field. For the case of the vortex slightly shifted from the pinning center the magnetic flux inside it is given by the following formula:

$$\frac{\Phi_1}{\Phi_0} = 1 + \frac{B_e \lambda^2}{\Phi_0}\left[\begin{array}{l} \arcsin\frac{d}{2\lambda} + \frac{d}{2\lambda}\sqrt{1-\left(\frac{d}{2\lambda}\right)^2} \\ -\frac{d}{\lambda}\cdot\frac{x}{\lambda} \end{array}\right] \tag{26}$$

while for deflection of the vortex core from the pinning center larger than $x_{c1}$ the magnetic flux in the vortex area is equal to:

$$\frac{\Phi_1}{\Phi_0} = 1 + \frac{B_e \lambda^2}{\Phi_0}\left[\frac{\pi}{2} - \arcsin\frac{x}{\lambda} - \frac{x}{\lambda}\sqrt{1-\left(\frac{x}{\lambda}\right)^2}\right] \tag{27}$$

Parameter $x_{c1}$ denotes the limiting value of the vortex deflection distinguishing these both regions, for which Eq. 24 or Eq. 25 are valid respectively and is described by the relation:

$$x_{c1} = \lambda\sqrt{1-\left(\frac{d}{2\lambda}\right)^2} \tag{28}$$

As follows from Eq. 27, for fully released vortex, it is shifted on the length x = λ, magnetic flux in vortex reaches the magnitude of $\Phi_0$. Magnetic flux $\Phi_1$ penetrating vortex determines then according to Eqs. 23-24 the maximal magnetic induction in the center of vortex core ***B(0)*** and influences in this way magnetic induction distribution inside the vortex, according to the relation 22. Conditions 22-24 determine new value of the vortex core radius, which creates normal state part of the vortex, in the language of Fig. 18 described by the effective coherence length. The new radius of the vortex core $\xi_1$ is determined according to Eqs. 25-27 by the relation:

$$\mu_0 H_c = \frac{\Phi_1}{2\pi\lambda^2}\exp(-\frac{\xi_1}{\lambda}) \tag{29}$$

where $H_C$ is thermodynamic critical magnetic field, $\mu_0$ magnetic permeability, while magnetic flux $\Phi_1$ is described by Eqs. 25-27. We apply then the condition joining the thermodynamic critical magnetic field with coherence length of an isolated vortex:

$$\mu_0 H_c = \frac{\Phi_0}{2\pi\lambda^2}\exp(-\frac{\xi_0}{\lambda}) \tag{30}$$

where $\Phi_0$ and $\xi_0$ describe the flux quantum and normal core radius of an isolated vortex. According to Eqs. 29 - 30 the expression for modified length of the normal core radius is given by the relation:

$$\frac{\xi_1}{\xi_0} = 1 + \kappa \ln \frac{\Phi_1}{\Phi_0} \tag{31}$$

where Ginzburg-Landau theory parameter $\kappa = \frac{\lambda}{\xi_0}$ characterizes here the kind of the superconductor. Present approach taking into account in the first approximation the electromagnetic nature of the interaction pancake vortex with the nano-sized pinning centers, leads therefore to the modification of the normal state core radius $\xi_1$ of the captured vortex. This parameter has essential meaning for an analysis of the energy barrier in the flux creep process, as it was shown in point 2.

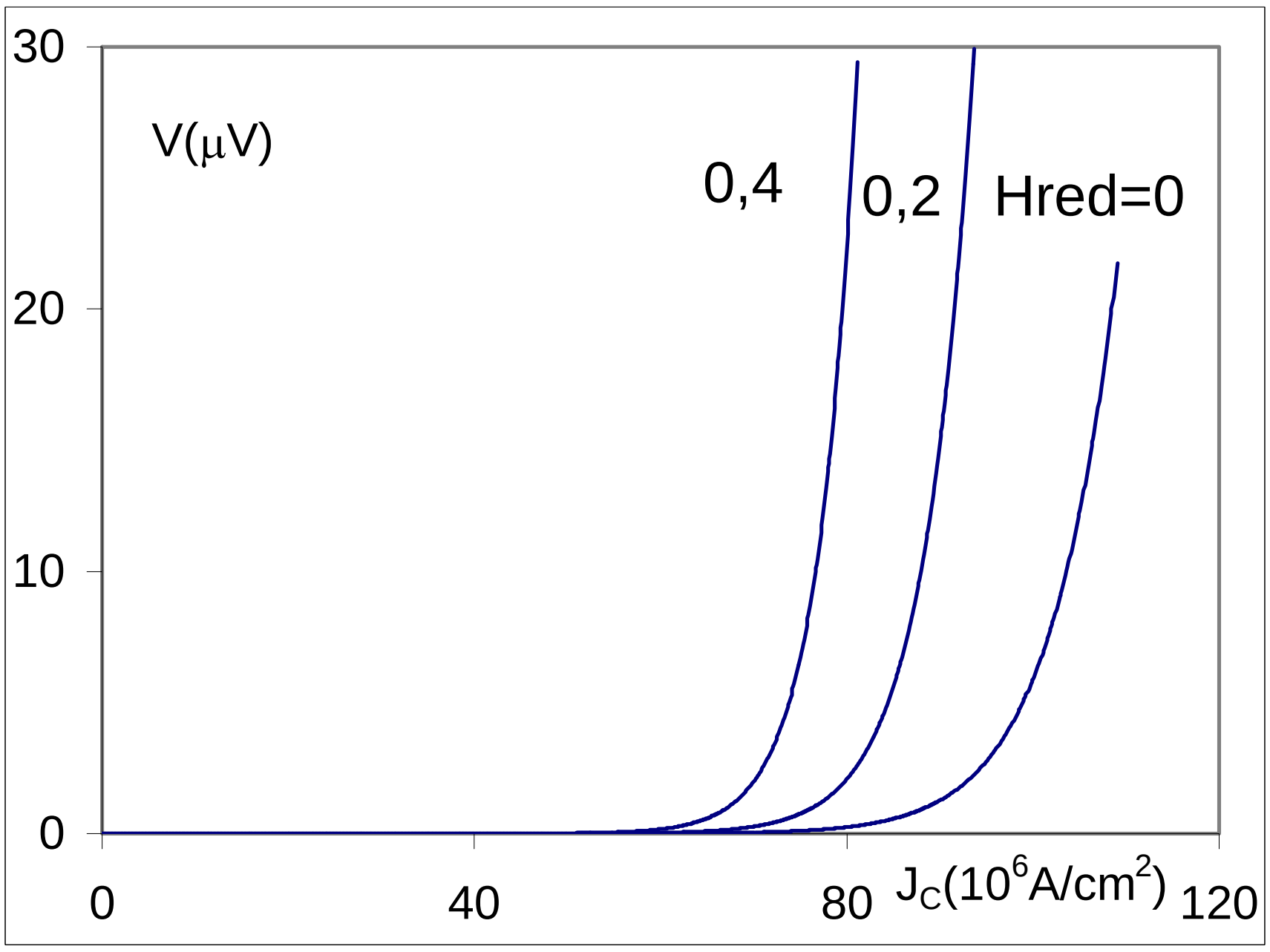

Fig. 19. Calculated current-voltage characteristics for the HTc superconductor in the function of the reduced applied magnetic field $H_{red}$

The results of the calculations current-voltage characteristics according to elaborated model in the function of the reduced magnetic field $H_{red} = \frac{B_e \lambda^2}{\Phi_0}$ are shown in Fig 19. Strong influence of the value $\boldsymbol{H_{red}}$ on the current-voltage characteristics is well seen in Fig. 19. Above result is in agreement with an experimental observation, that applied magnetic field decreases the critical current. Critical current density is determined then, as the value for which potential barrier vanishes. In Fig 20 is presented the influence on the critical current density of the parameter $\alpha$, describing the rigidity of the vortex lattice, also in the function of the magnetic flux $\Phi_1$ appearing in the relations 25-27.

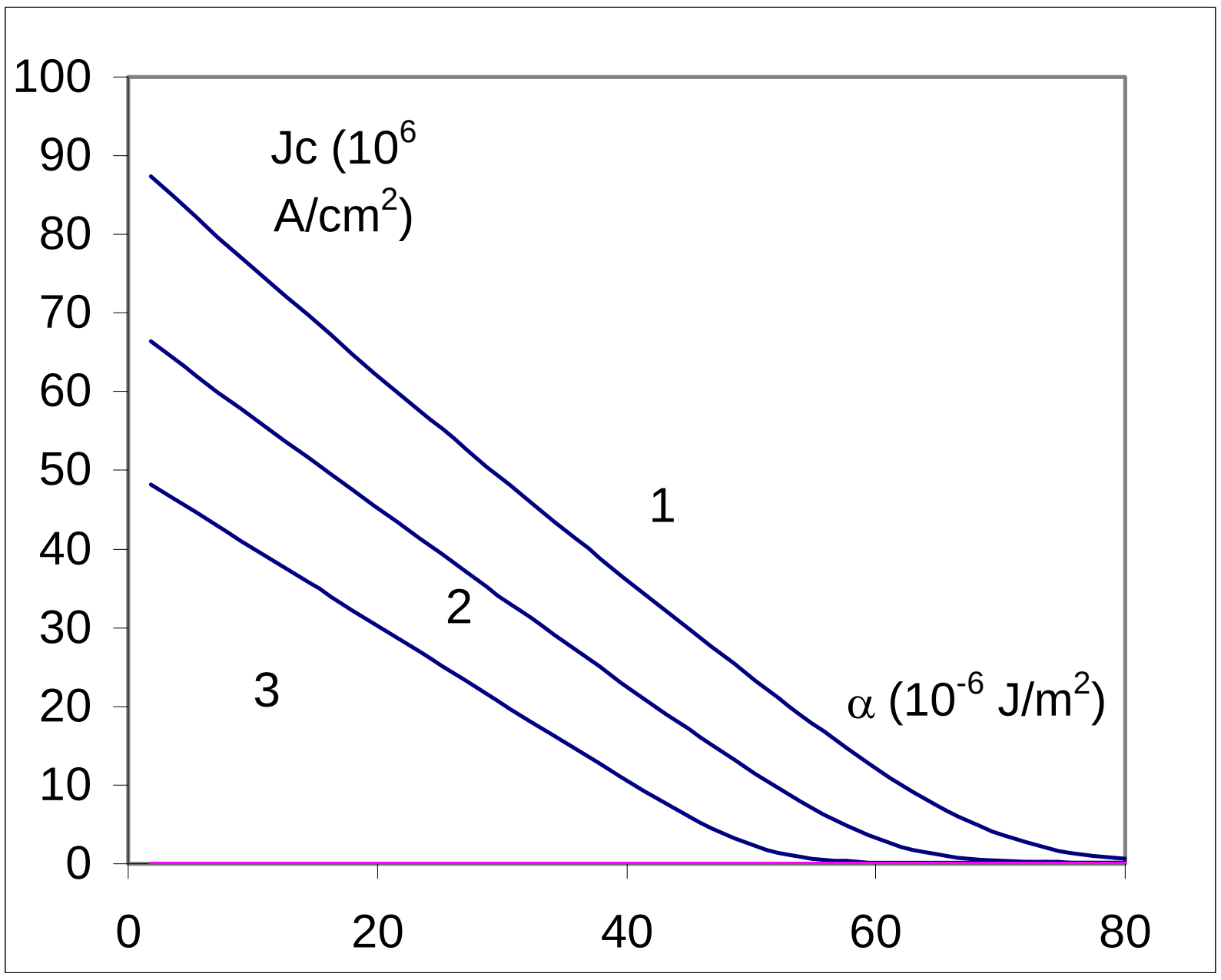

Fig. 20. The dependence of the critical current on the rigidity of the vortex lattice, expressed by the parameter ***α*** in the function of the magnetic flux $\Phi_1$: (1) $\Phi_1 = \Phi_0$ (2) $\Phi_1 > \Phi_0$ (3) $\Phi_2 > \Phi_1$

New model of the pinning mechanism allows therefore to predict the critical current of the HTc materials in function of the vortex capturing parameters. Critical current is essential parameter from the point of view of the application HTc materials in the energy transport process. From other side critical current determines too the screening properties of bulk HTc superconductors, which are important at application of these materials as permanent magnet, for instance in levitation process. We should notice that in bulk HTc materials, due to their granular structure, there appear not only intragranular currents but also intergarnular, which can be treated as Josephson's like currents. The peculiar screening properties of HTc superconductors are related to the flux trapping, while analysis of this effect is performed in the next point.

## 6. Trapped magnetic flux analysis in HTc ceramics

Trapped magnetic flux (Ftr) is important parameter dependent on the flux pinning and critical current. It is defined as the remanent moment of the magnetization curve in the magnetic field cycle $0 \rightarrow B_m \rightarrow 0$, where $B_m$ is maximal magnetic induction in the cycle. An example of an irreversible magnetization curves for high temperature superconductor (Bi-2223 bulk ceramic) and low temperature superconductor (semiconducting lanthanum selenide) is shown in Figs. 21-22. Both these curves indicate on the existence of magnetic irreversibility, which means that trapped flux is common property of type II superconductors. The existence of small flux jumps is also observed here for both directions of the magnetic field variation.

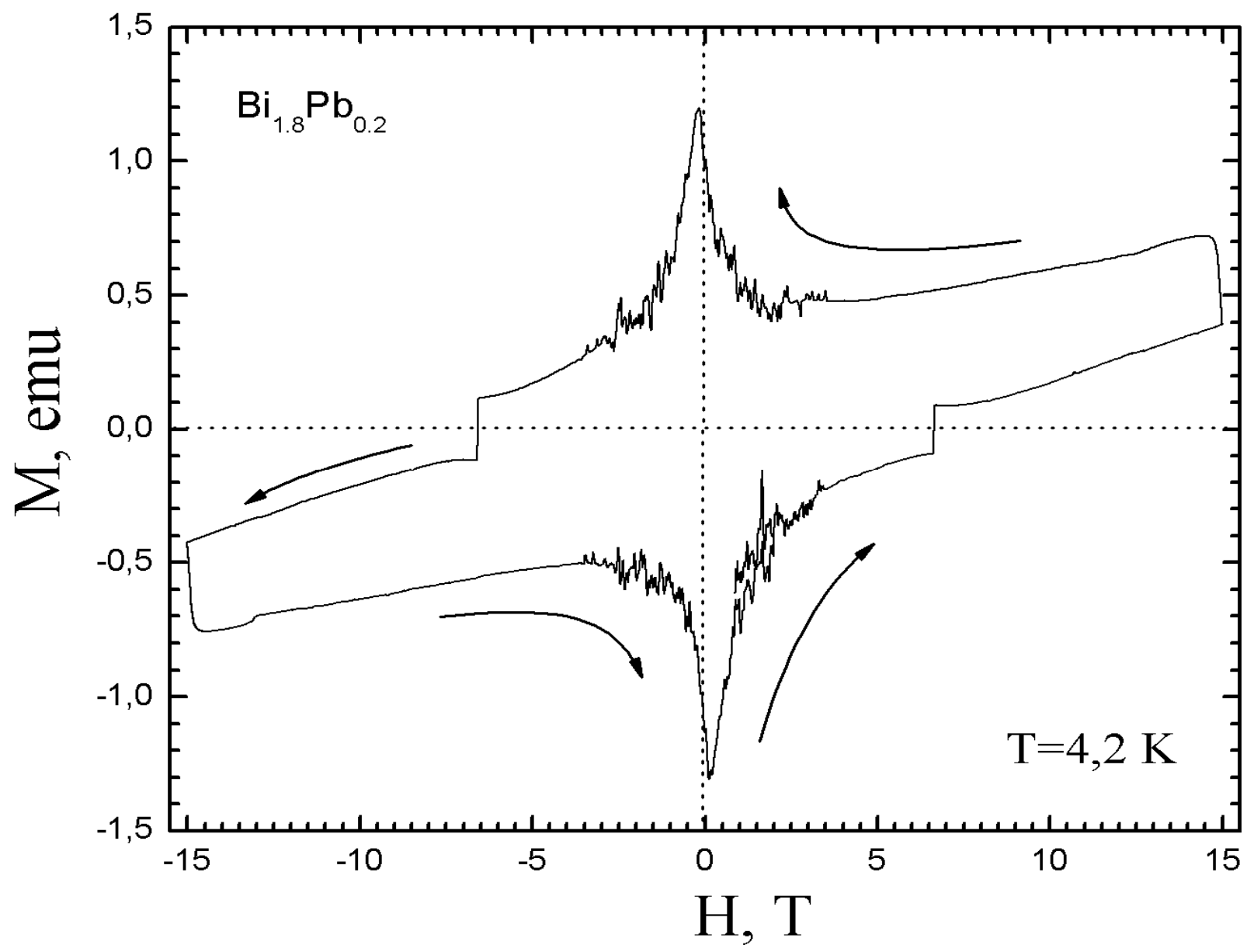

Fig. 21. Full magnetic hysteresis loop of the $Bi_{1.8}Pb_{0.2}Sr_2Ca_2Cu_3O_8$ HTc superconducting ceramic in high magnetic field

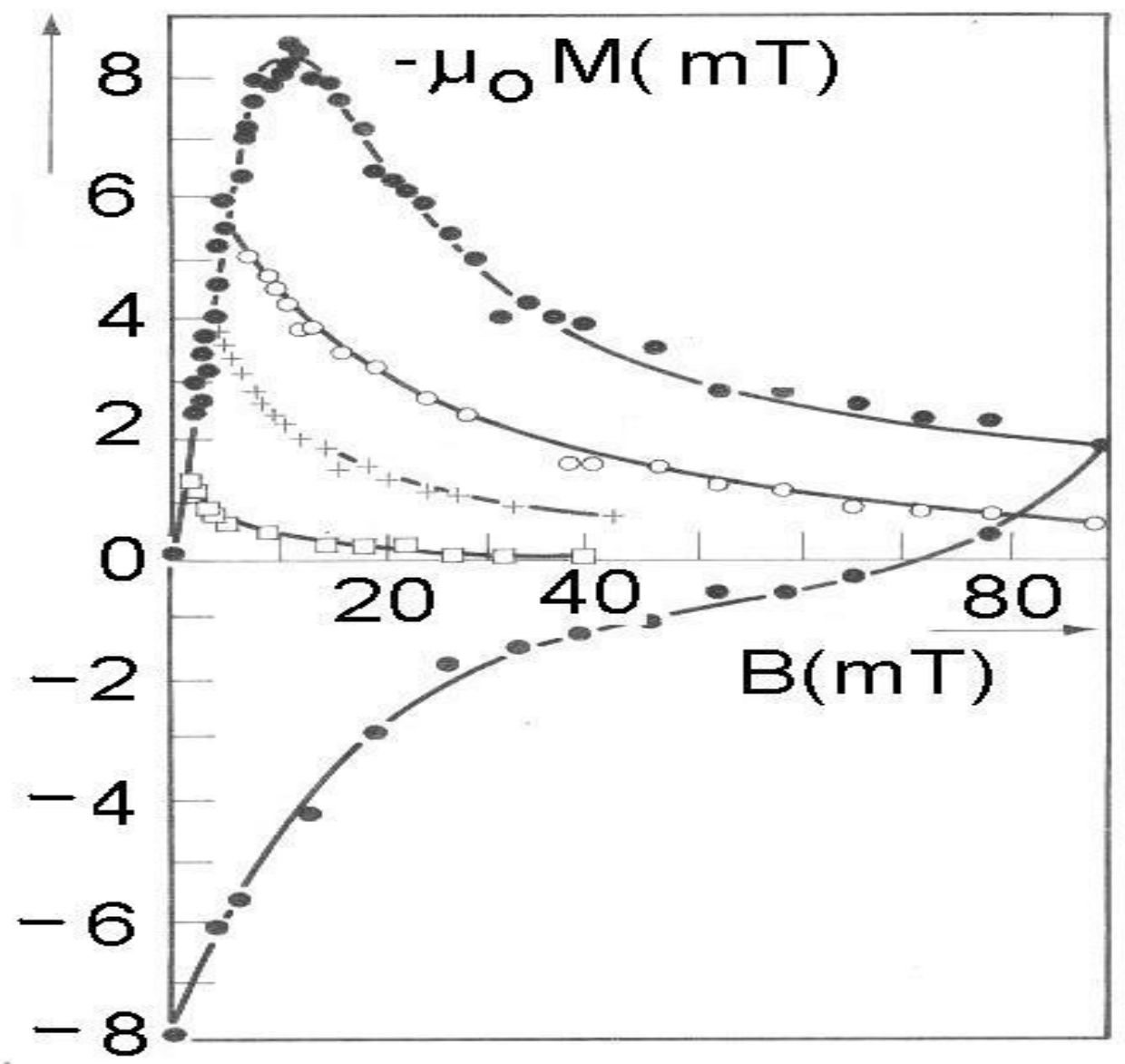

Fig. 22. Magnetic irreversible curves of the four samples of the $La_{3-x}V_xSe_4$ semiconducting superconductor, with increasing electron carrier concentration. V – lanthanum vacancy.

Trapped magnetic flux is measured by the Hall probe method as well as just by the remanent moment of the magnetization. In best macro-granular YBaCuO samples trapped flux reaches 17 T at 29 K, which value clearly indicates on the magnitude of this effect in HTc superconductors and makes therefore the attractive possibility of applying these materials as the permanent magnets. Fig. 23 presents the calculated model of levitating train with using permanent magnets oppositely oriented and levitating above them HTc superconductor in Meissner state. Similar experiment can be performed for the superconductor in the flux trapped state.

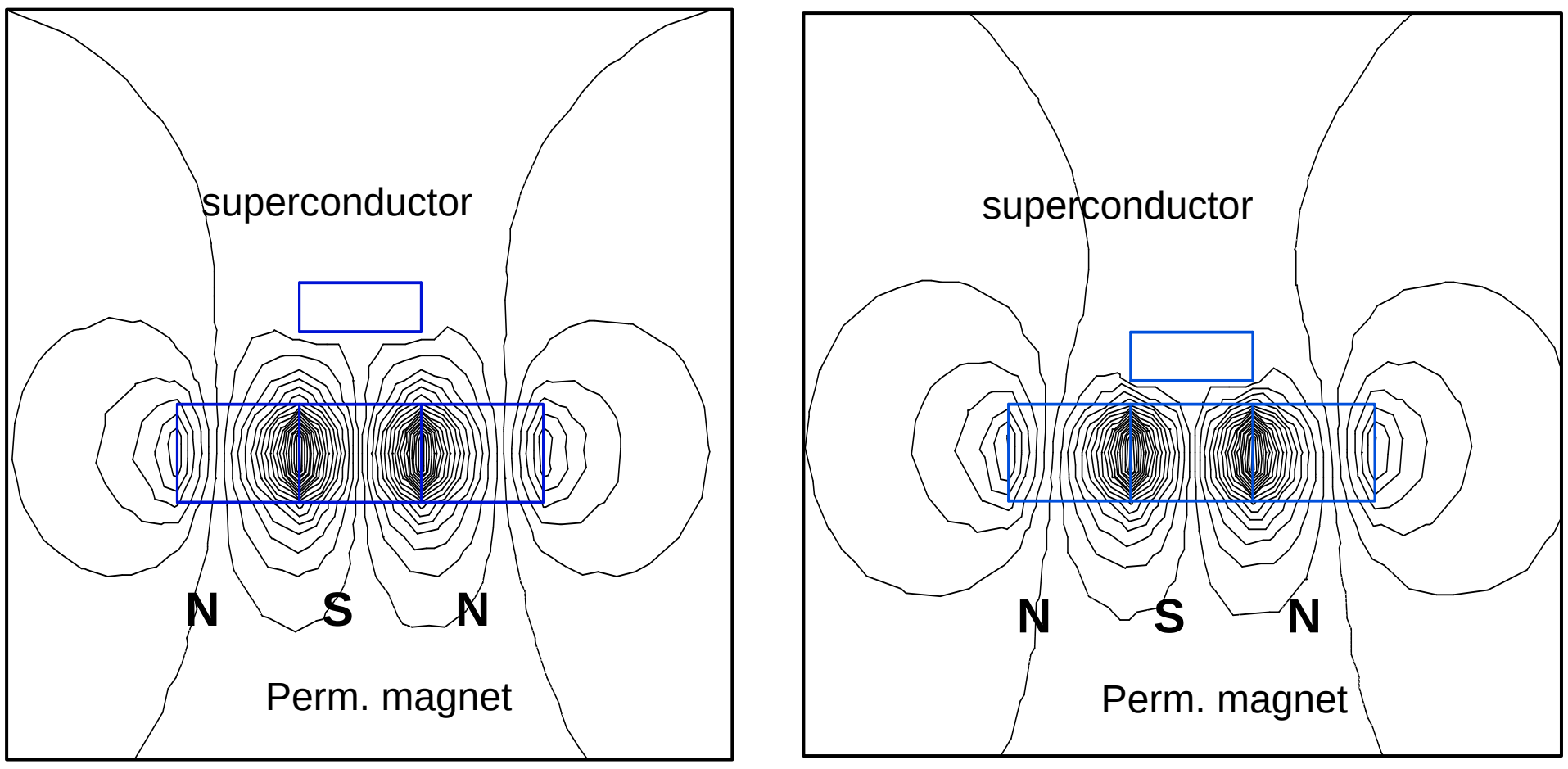

Fig. 23. Calculated magnetic field lines in the model of the levitating train - superconductor above two oppositely oriented permanent magnets, for two distances

Fig. 24 shows magnetic field profiles in the superconducting bearing constructed from HTc superconducting tube rotating on the core composed from six permanent magnets.

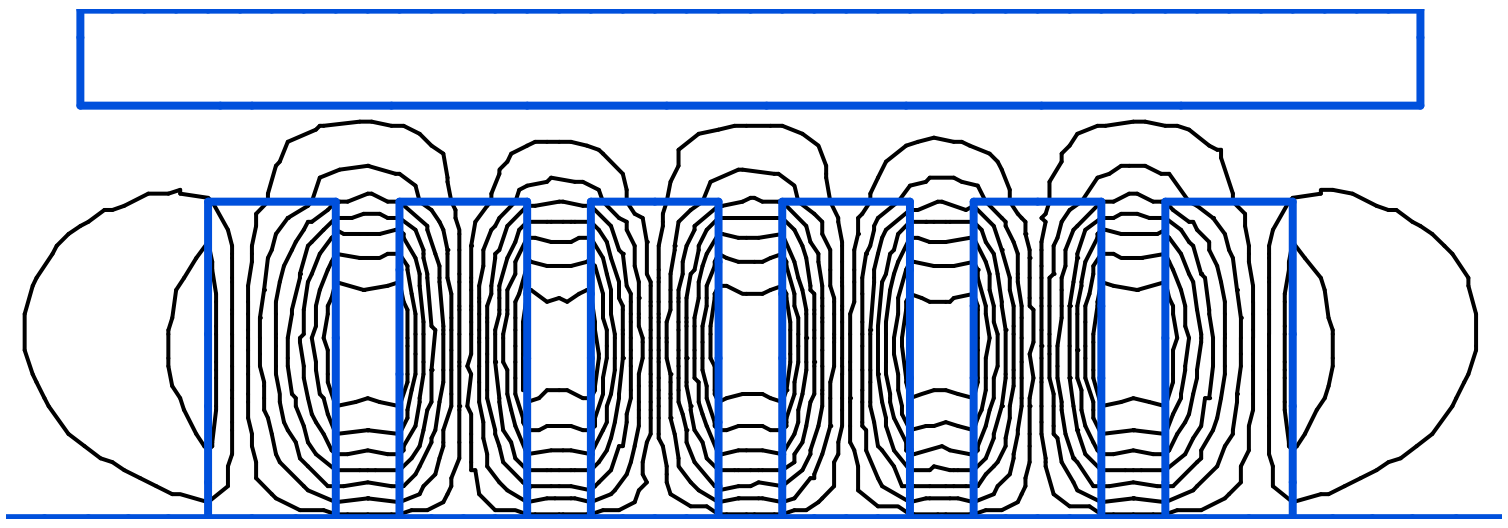

Fig. 24. Cross-section of the magnetic bearing composed from the HTc superconducting bearing levitating above 6 permanent magnets.

The scheme of the magnetic induction distribution in the flux trapped state presents Fig. 25. Existence of the intergranular Josephson's currents, mentioned previously and the current inside grains lead to the tooth-like shape of the magnetic induction profile given schematically here.

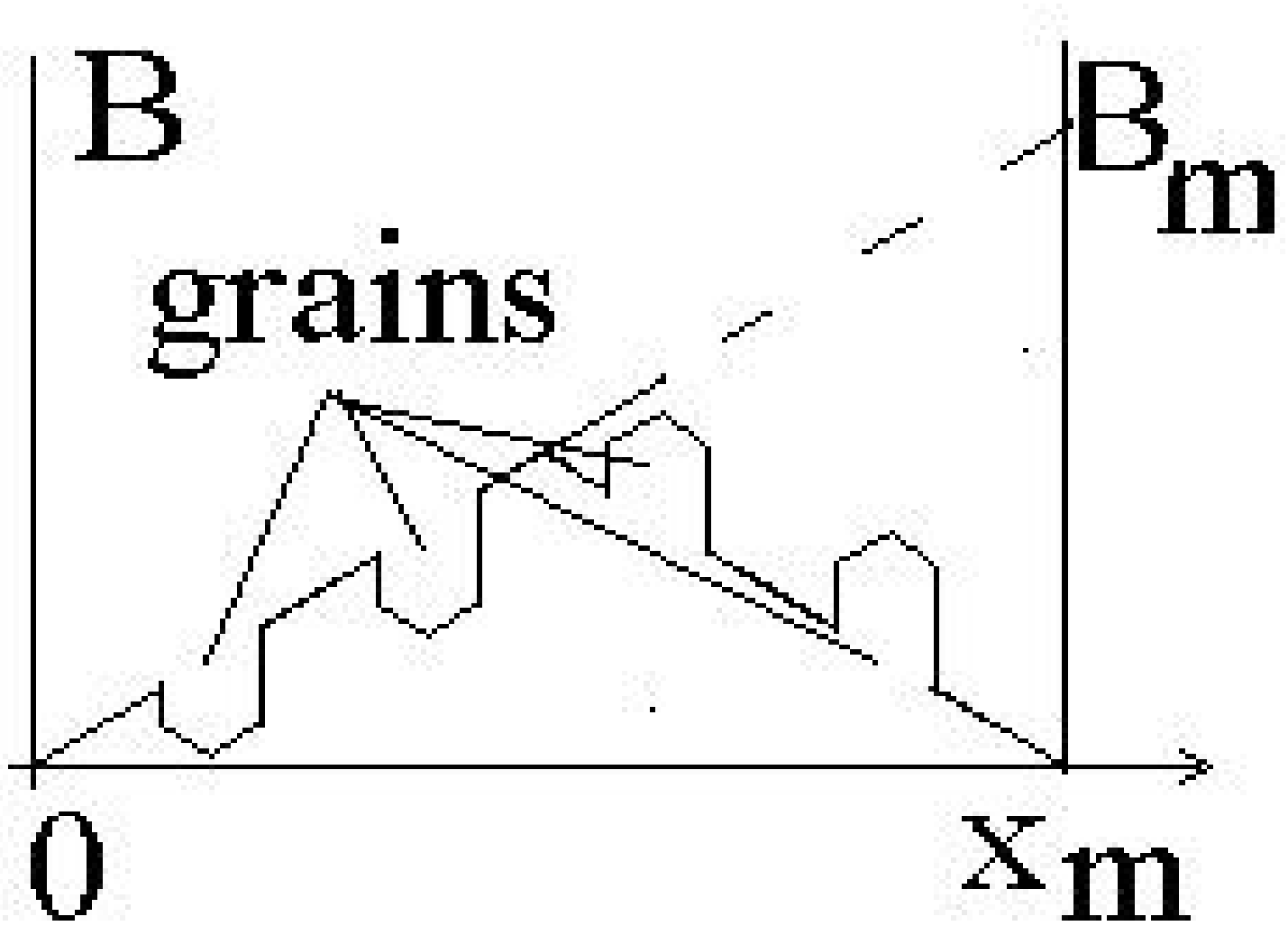

Fig. 25. Magnetic induction profiles in the flux trapped state of the HTc ceramic superconductor for the magnetic field cycle 0 – $B_m$ - 0. Sharp changes of the magnetic profiles are due to existence of the grains.

At an aim of an analysis of the trapped flux value we have considered four cases dependent on the amplitude of the external magnetic induction $B_m$ in the magnetizing cycle. First case has the place for the maximal induction $B_m$ smaller than the first penetration field $B_{c1}$. Then the magnetic induction does not penetrate into the volume of the superconductor, while trapped flux $F_{tr}$ vanishes therefore. In the second region of magnetic field variations described by the relation: $\xi > B_m^1 > 0$, where $B_m^1 = B_m - B_{c1}$, trapped flux being mathematically the integral over the superconductor volume from the induction profile shown in the Fig. 25 is expressed by the following relation, in an approximation of the constant current density in the cylindrical geometry sample:

$$F_{tr} = \frac{B_m^{1^2}}{2\xi^2}\left[\xi - \frac{B_m^1}{2} + nB_{sg}\right] \qquad (32)$$

Parameter $\xi$ describes here full penetration magnetic induction inside the superconducting material of the radius R, without the surface barrier effects:

$$\xi = \mu_0 j_c R \qquad (33)$$

In Eq. 32 has been taken into account according to the Fig. 25 the existence of the superconducting grains immersed into intergranular matrix, filling of which is described by the relative concentration n, varying between 0 and 1. An amount of the trapped flux connected with an individual grain, on its cross-section we denote by the symbol $B_{sg}$, described by the dependence:

$$B_{sg} = Bc_{1g} + \frac{\xi g}{3} \qquad (34)$$

$\boldsymbol{Bc_{1g}}$ denotes first critical magnetic field in the superconducting grains, while $\boldsymbol{\xi_g}$ is determined by the relation:

$$\xi g = \mu_0 j_{cg} R_g \qquad (35)$$

$\boldsymbol{R_g}$ is the radius of the superconducting grain and $\boldsymbol{j_{cg}}$ intragrain current density.

For the next range of the magnetic induction increase determined by the condition $2\xi \geq B_e^1 \geq \xi$ trapped magnetic flux magnitude, averaged to the cylindrical sample cross-section is equal:

$$F_{tr} = \frac{2nB_{\lambda g}}{\xi^2}\left(B_e^1\xi - \frac{\xi^2}{2} - \frac{B_e^1}{4}\right) + \frac{2\xi}{3}\left(\frac{1}{2} - \left(1 - \frac{B_e^1}{2\xi}\right)^3\right) \qquad (36)$$

For maximal applied magnetic induction in the field cycle, described by the condition $B_e^1 \geq 2\xi$ trapped flux reaches the constant value:

$$F_{tr} = \frac{\xi}{3} + nB_{1g} \quad (37)$$

Theoretical results given by Eqs. 32-37 allow then to determine influence of various material parameters on the flux trapping, which permits then to find the optimal value of these parameters. Examples of the calculations of the dependence of the trapping magnetic flux in the cylindrical sample on the maximal magnetic induction amplitude $B_m$, versus materials parameters are given in the Figs. 26 - 29 and really indicate on an importance of given parameter for receiving maximal flux trapping.

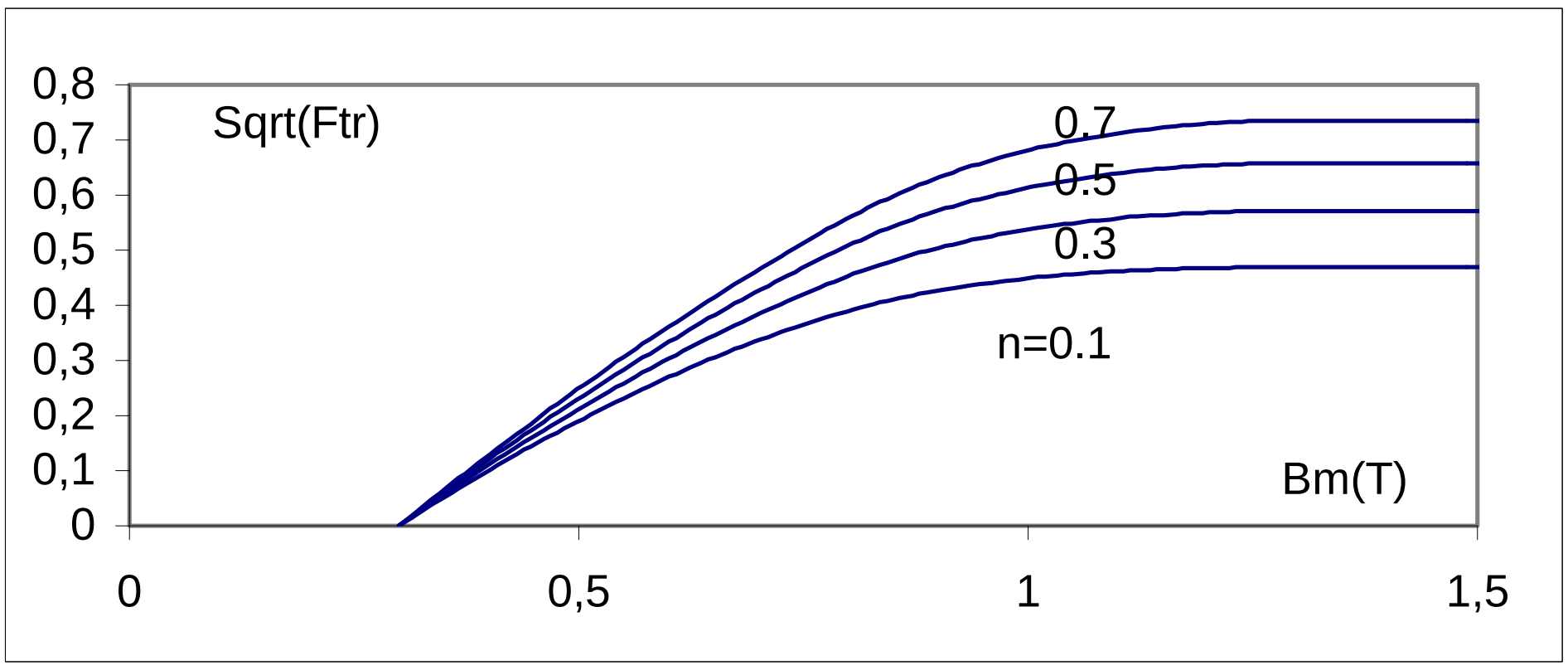

Fig. 26. Influence of the fulfillment with the superconducting grains of the superconductor on the square root from the flux trapping Ftr. Numbers at curves indicate the values of the reduced grain concentration varying between 0<n<1.

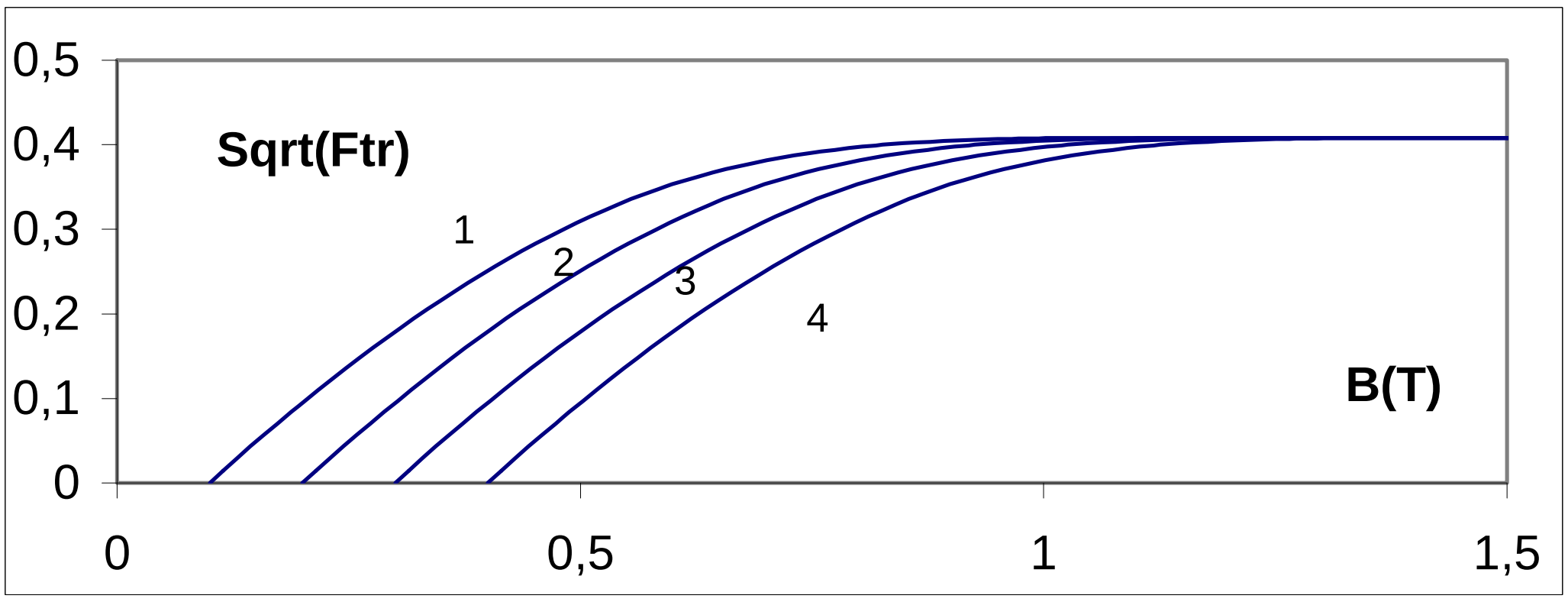

Fig. 27. Influence of the first critical magnetic field $B_{c1}$ of the superconducting matrix on the trapped magnetic flux Ftr: (1) $B_{c1}$ = 100 mT, (2) 200 mT, (3) 300 mT, (4) 400 mT.

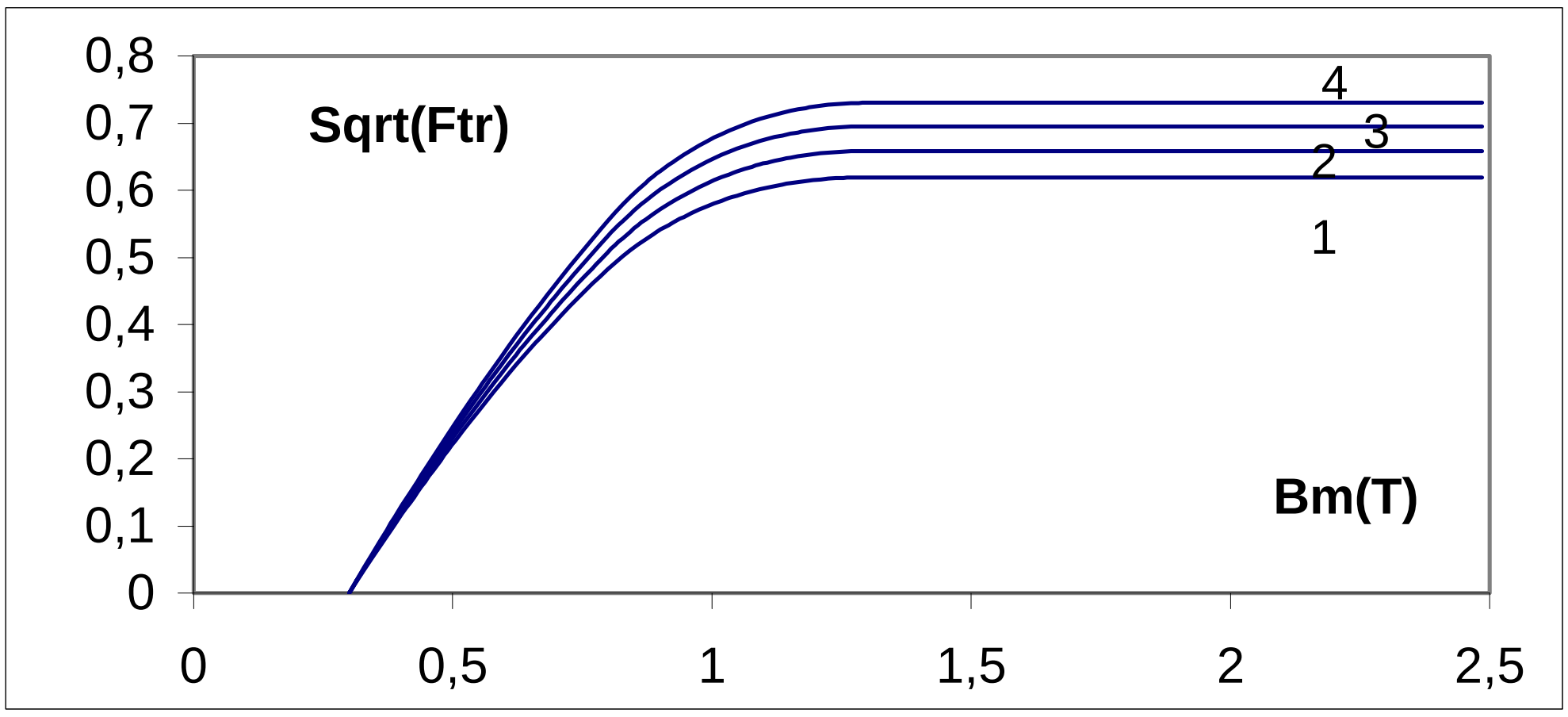

Fig. 28. Influence of the first critical magnetic field $B_{c1q}$ of the superconducting grains on the trapped magnetic flux Ftr: (1) $B_{c1q}$ = 100 mT, (2) 200 mT, (3) 300 mT, (4) 400 mT.

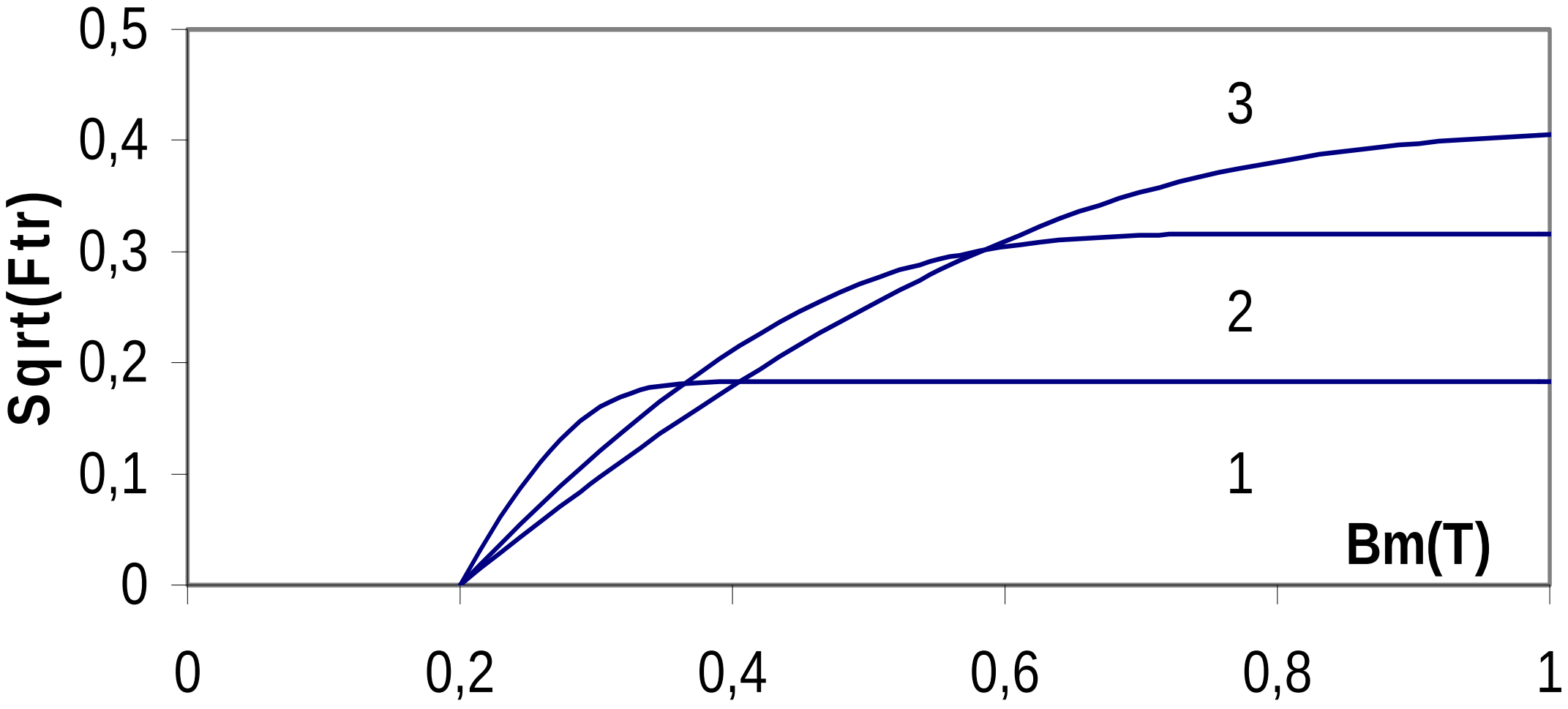

Fig. 29. The influence of the intergranular critical current of the ceramic superconductor on square root from flux trapped (Sqrt(Ftr)) versus maximal magnetic field in the cycle. Indexes 1-3 indicate the case of the increasing critical current.

Beside cylindrical samples described above significant is too the case of the flat ones, which roughly corresponds to the technical tapes and plates. We will analyze now therefore flux trapping for that geometry. As previously four cases dependent on an external magnetic induction amplitude have been considered. The results in each of them are described by an Eqs. 38-41, while graphical presentation of obtained relations is shown in Figs. 30-33.

$$F_{tr} = 0 \tag{38}$$

$$F_{tr} = \frac{B_m^2}{4B_p} \tag{39}$$

$$F_{tr} = B_m - \frac{B_m^2}{4B_p} - \frac{B_p}{2} + nB_g\left(\frac{B_m}{B_p} - 1\right) \tag{40}$$

$$F_{tr} = \frac{B_p}{2} + nB_g \tag{41}$$

$B_m$ is maximal magnetic induction in field cycle, while $B_p$ induction of the full penetration.

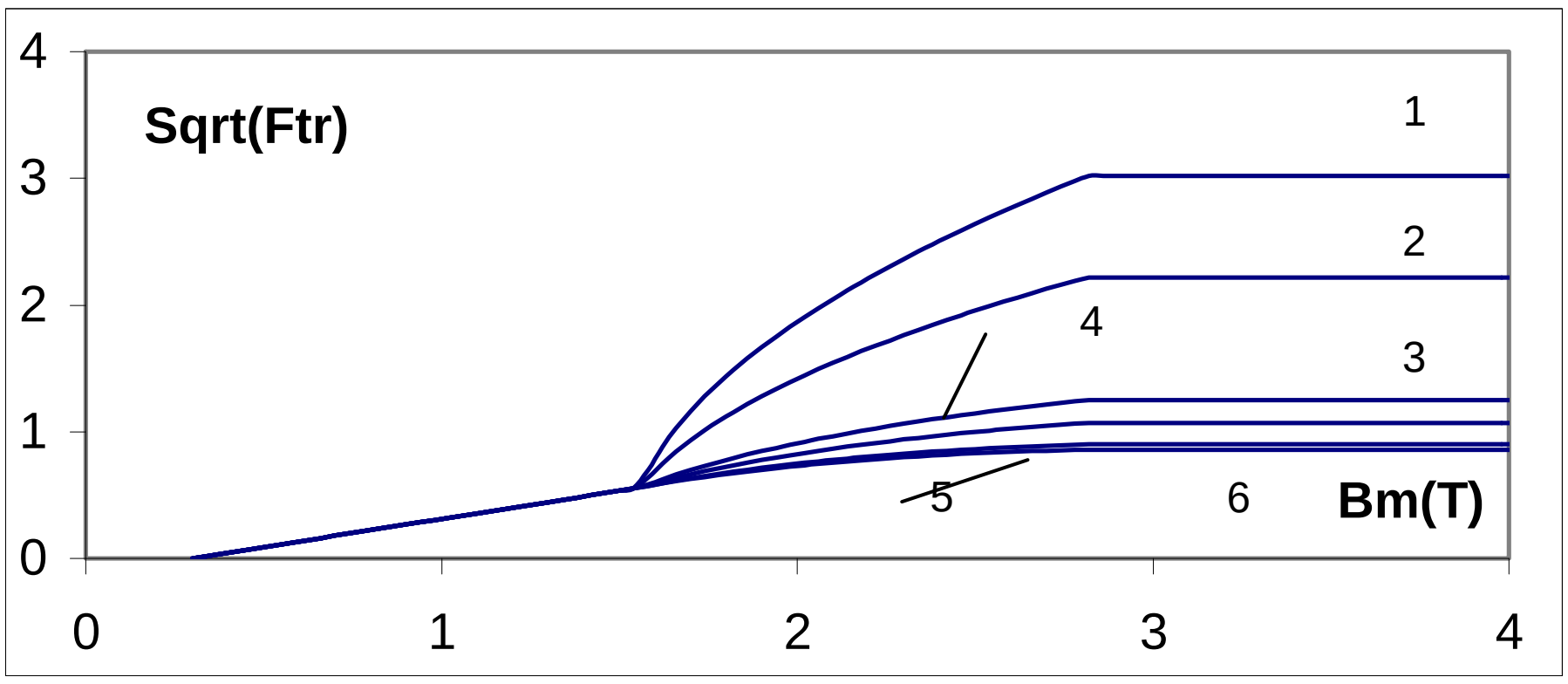

Fig. 30. Dependence of the flux trapping in superconducting plate on maximal magnetic induction in the cycle for various critical current density inside the superconducting grains. Subsequent curves refer to the following current densities: (1) $j_{cq}= 10^{12}$, (2) $5*10^{11}$, (3) $10^{11}$, (4) $5*10^{10}$, (5) $10^{10}$, (6) $5* 10^9$ A/$m^2$

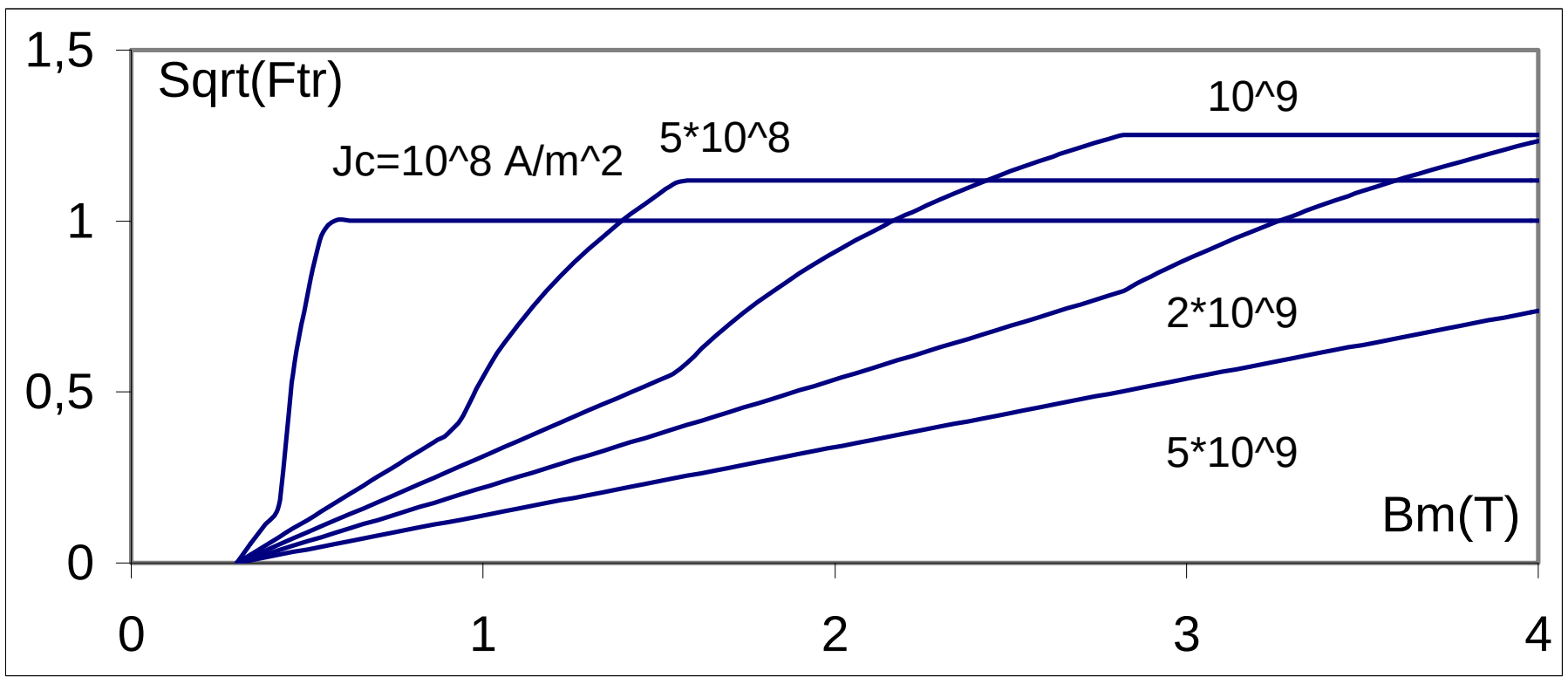

Fig. 31. Dependence of the square root of the flux trapping in superconducting plate on maximal magnetic induction in the cycle Ftr for various critical current density inside the superconducting matrix.

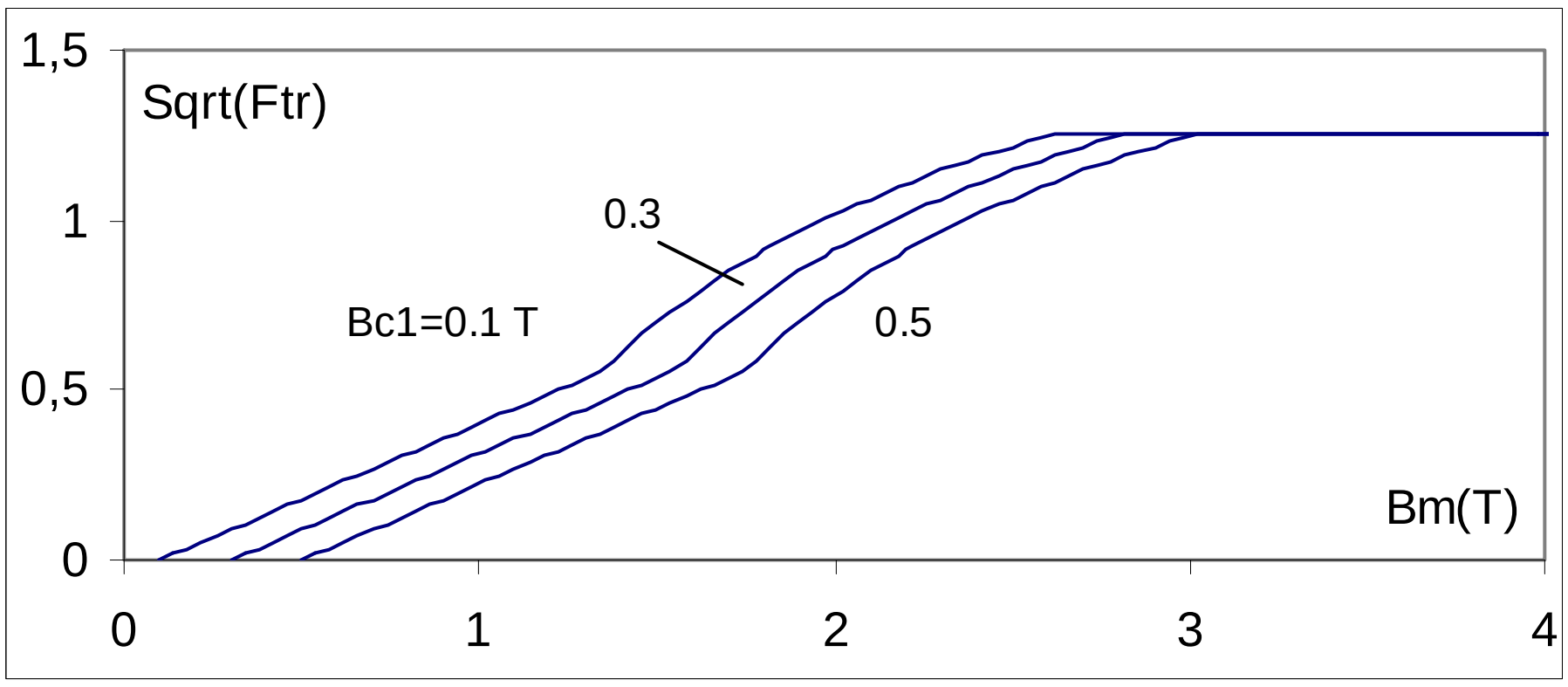

Fig. 32. Dependence of the flux trapping in superconducting plate on maximal magnetic induction in the cycle for various first penetration field Bc1.

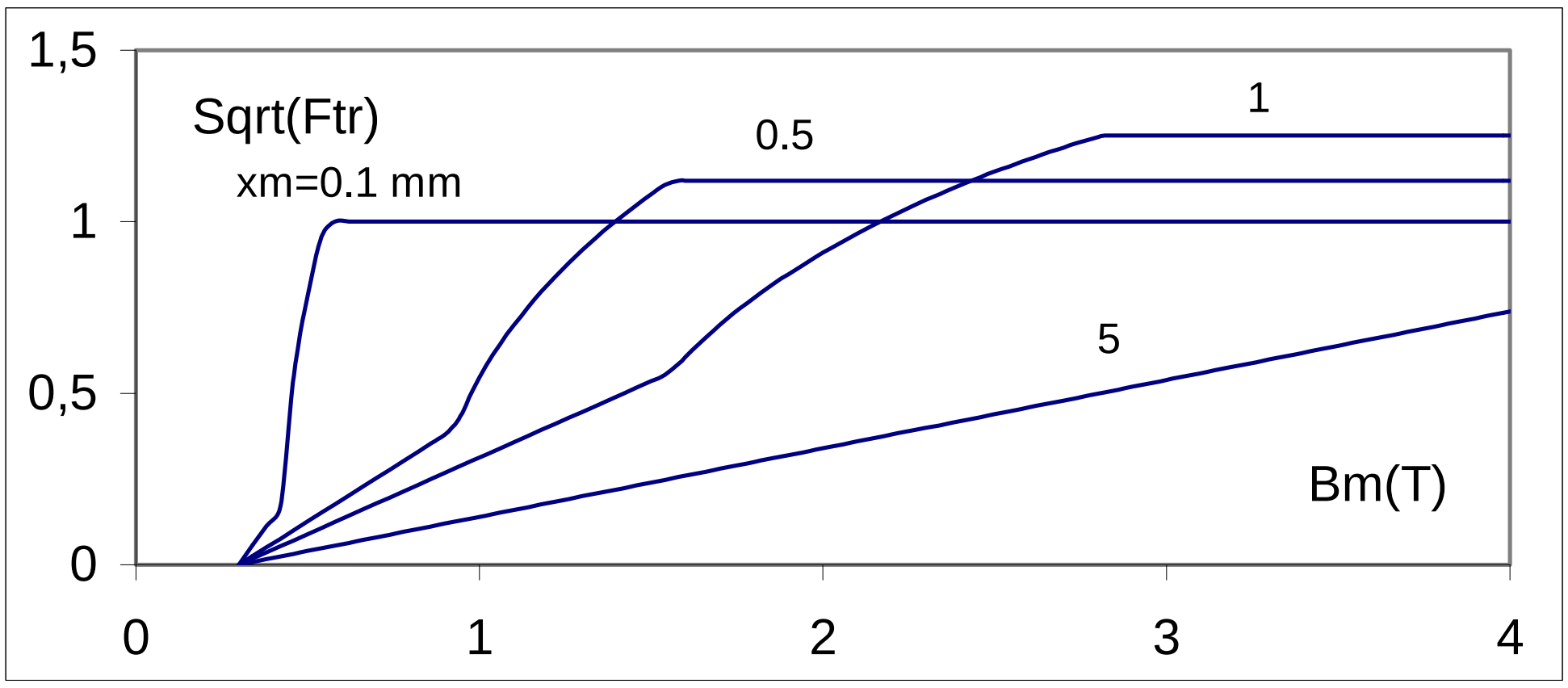

Fig. 33. Dependence of the flux trapping in superconducting plate on maximal magnetic induction in the cycle for various flat sample thickness $2x_m$ .

Presented mathematical model permits also for taking into account the surface barrier arising on the smooth edge of the superconductor. Then we obtain the following set of equations describing flux trapping in the function of the maximal external induction amplitude B defined now according to:

$$B = B_{max} - B_{cl} - \Delta B \tag{42}$$

In Eq. 42 $B_{c1}$ is first critical induction, while $\Delta B$ just surface barrier. In the first range of field variation described by the relation:

$$B_{cl} + \Delta B \leq B_{max} \leq B_{cl} + 2\Delta B \tag{43}$$

trapped flux is equal to:

$$F_{tr} = \frac{j_c}{R^2}\left[\frac{\left(R + \frac{B}{j_c}\right)^3}{3} + R^2\left(\frac{B}{j_c} - \frac{R}{3}\right)\right] \tag{44}$$

Another relation has the place for magnetic inductions inside the field range:

$$B_{cl} + 2\Delta B \leq B_{max} \leq j_c R + B_{cl} + \Delta B \tag{45}$$

$$F_{tr} = \frac{1}{6R^2 j_c}\left[3R\left(B^2 - \Delta B^2 + 2\Delta B\right) - \frac{\Delta B^3 + 3B\left(B\Delta B + B^2 - \Delta B^2\right)}{2j_c}\right] \tag{46}$$

In the magnetic induction range described by condition:

$$B_{c1} + \Delta B + j_c R \leq B_{max} \leq B_{c1} + \Delta B + 2 j_c R \tag{47}$$

appears relation:

$$F_{tr} = \frac{j_c}{3R^2}\left[\frac{3R}{2j_c^{\,2}}\left(B^2 - \Delta B^2 + 2\Delta B\right) + \left(\frac{B}{j_c} - R\right)^2 - \frac{\Delta^3 + 3B\left(B\Delta B + B^2 - \Delta B^2\right)}{4j_c}\right] \tag{48}$$

For still higher magnetic field described by:

$$2j_c R + B_{c1} + \Delta B \le B_{\max} \tag{49}$$

trapped flux reaches constant value:

$$F_{tr} = \Delta B + \frac{j_c R}{3} \tag{50}$$

Presented model allows to follow the influence of superconducting ceramic material parameters on the flux trapping, what should be useful in technological works having as the purpose optimalize this flux and generally magnetic parameters of the superconductors. On the other hand in this way it is possible from the flux trapping measurements to receive information on such superconducting parameters as critical current, first penetration field, surface barrier. Experimental results of measurements flux trapping performed on YBaCuO ceramic Fe doped are shown in Fig. 34 and indicate on the good agreement with the theoretical predictions shown in Figs. 25-33, which independently confirms theoretical model.

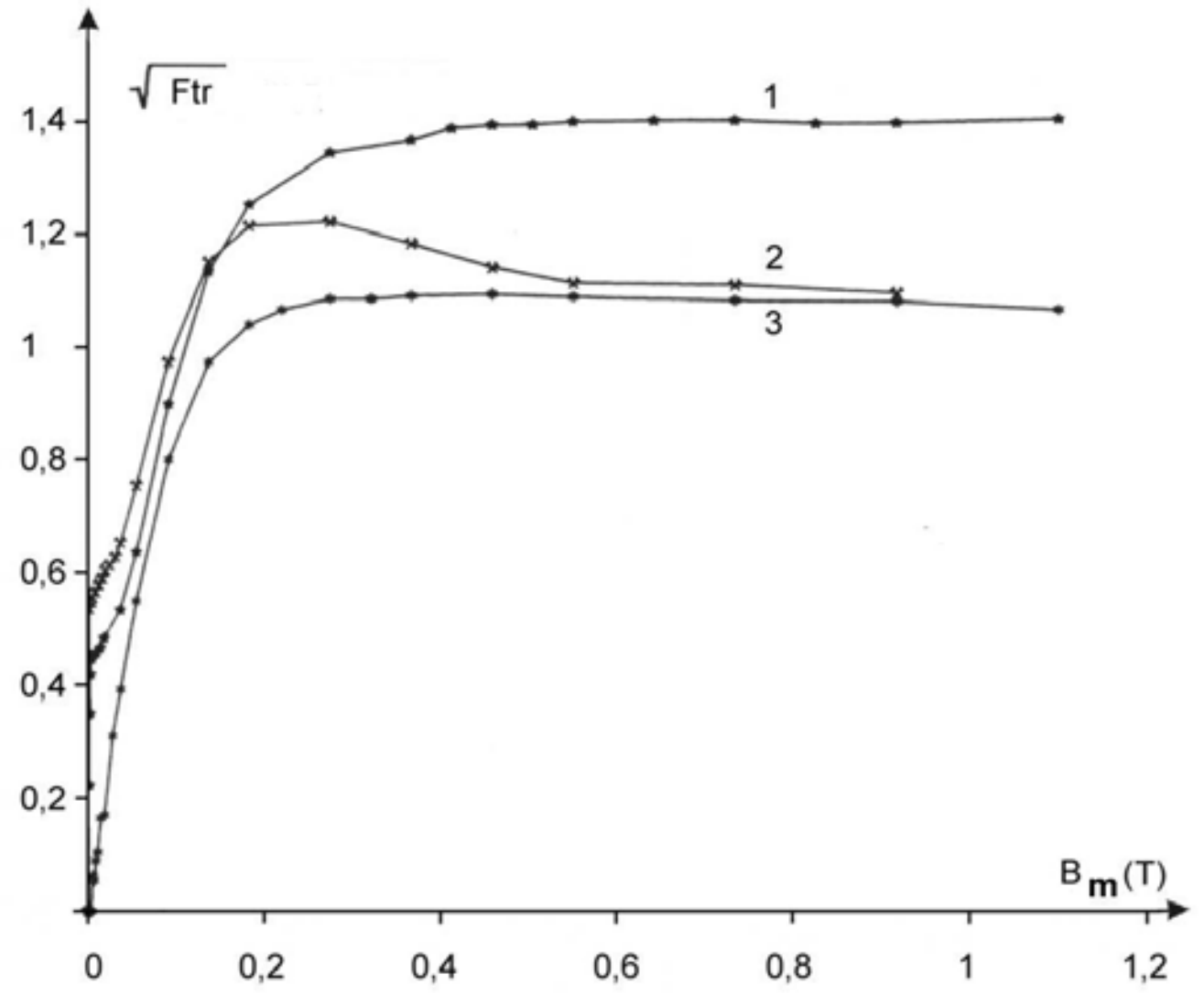

Fig. 34. Experimental dependence of the measured flux trapped versus maximum applied magnetic induction $B_m$ for HTc superconducting ceramic of the YBaCuO type doped with Fe impurity (according to the reference [8])

As it has been proved previously, defects and grains boundaries determine the critical current in HTc superconductors and their magnetic properties. Alternating current flow from other side leads to generation of losses in HTc tapes. Beside pure hysteresis type losses in the modern 2G tapes the a.c. losses are enhanced due to the magnetic nickel substrate on which HTc films of YBaCuO type are deposited. Magnetic substrate leads to an enhancement of the alternating magnetic field in the superconducting film. An example of the calculations influence on a.c. losses ***L*** of the magnetic characteristics of the nickel substrate described by an expression **B=0.65 tanh(β·H),** where ***β*** is material parameter is shown in Fig. 35. Presented here results indicate on important enhancement of a.c. losses with parameter ***β*** describing the shape of the magnetic material characteristics.

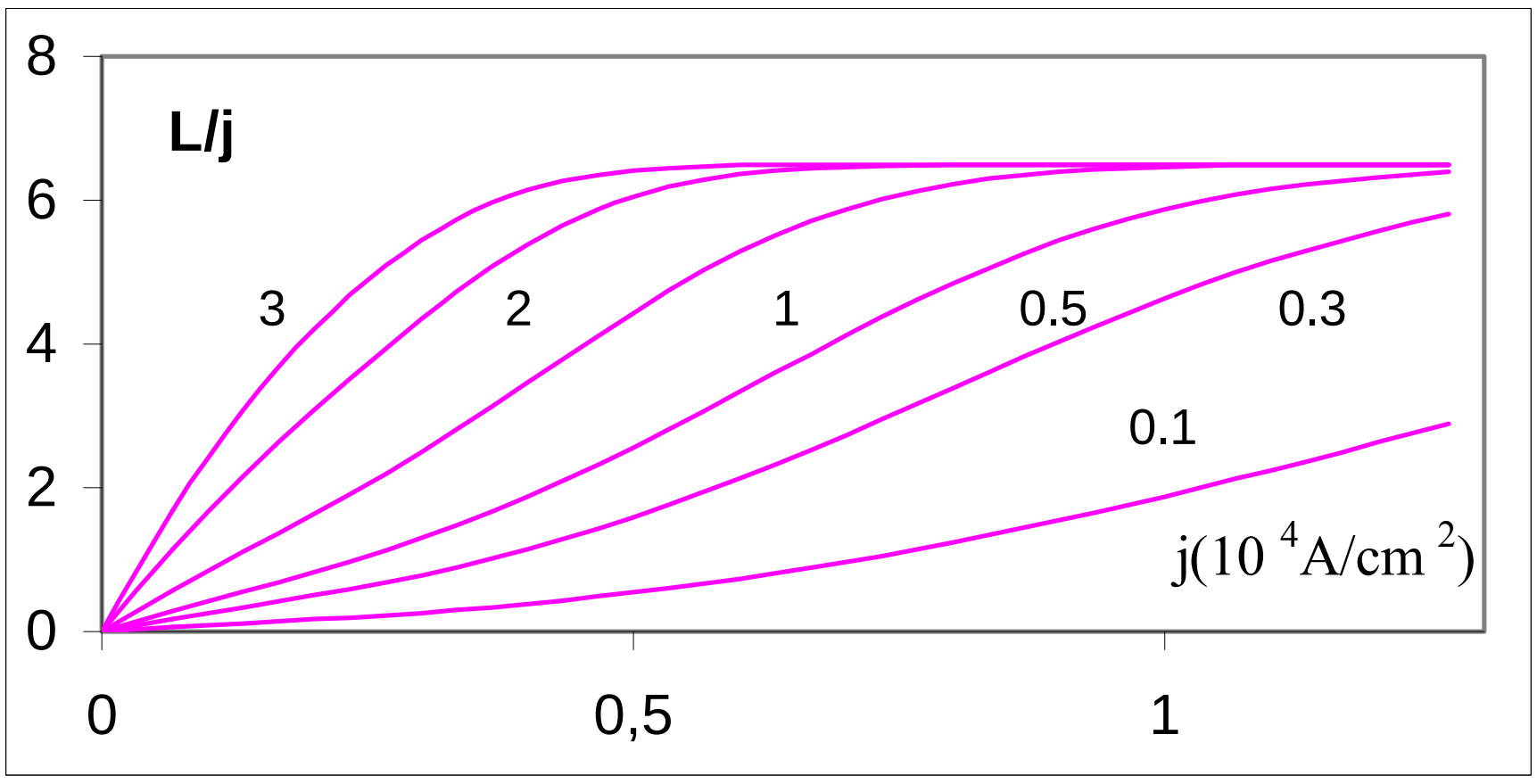

Fig. 35. Calculated a.c. losses in reduced form ***L/j*** in superconducting tape of the second generation 2G, normalised to the transport current density ***j*** versus this current density, for various values of the parameter ***β*** given at each curve

## 7. Conclusions

New model of the pinning interaction has been presented, allowing determine current-voltage characteristics, critical current, flux pinning and flux trapping in the HTc superconductors. Model is based on the detailed analysis of the interaction pancake type vortices with the nano-sized pinning centers. Comparison with experimental data of the theoretical results is given and good agreement has been received. Influence of the critical current on a.c. losses in tapes of 2G type with magnetic substrate is discussed briefly.

## 8. Acknowledgments

Author is grateful to MSc. Eng. Joanna Warycha from Electrotechnical Institute, Wroclaw Department (Poland) for performing scanning electron microscopy analysis.